\documentclass[showpacs,preprintnumbers]{revtex4}
\usepackage{amsmath}
\usepackage{graphicx}
\usepackage{amsfonts}
\usepackage{amssymb}
\begin{document}
\textbf{The exact proof that Maxwell equations with the 3D E}

\textbf{and B are not Lorentz covariant equations. The new }

\textbf{Lorentz invariant field equations}\bigskip\bigskip

\qquad Tomislav Ivezi\'{c}

\qquad\textit{Ru\mbox
{\it{d}\hspace{-.15em}\rule[1.25ex]{.2em}{.04ex}\hspace{-.05em}}er Bo\v
{s}kovi\'{c} Institute, P.O.B. 180, 10002 }

\qquad\textit{Zagreb, Croatia}

\textit{\qquad ivezic@irb.hr\bigskip\medskip}

In this paper it will be exactly proved both in the geometric algebra and
tensor formalisms that the usual Maxwell equations with the three-dimensional
(3D) vectors of the electric and magnetic fields, $\mathbf{E}$ and
$\mathbf{B}$ respectively, are not, contrary to the general opinion, Lorentz
covariant equations. Consequently they are not equivalent to the field
equations with the observer independent quantities, the electromagnetic field
tensor $F^{ab}$ (tensor formalism) or with the bivector field $F$ (the
geometric algebra formalism). Different 4D algebric objects are used to
represent the standard observer dependent and the new observer independent
electric and magnetic fields. The proof of a fundamental disagreement between
the standard electromagnetism and the special relativity does not depend on
the character of the 4D algebric objects used to represent the electric and
magnetic fields. The Lorentz invariant field equations are presented with
1-vectors $E$\ and $B$, bivectors $E_{HL}$ and $B_{HL}$\ and the abstract
tensors, the 4-vectors $E^{a}$\textbf{\ }and\textbf{\ }$B^{a}$. All these
quantities are defined without reference frames. Such field equations are in a
complete agreement with experiments.\bigskip

\bigskip\medskip

\textbf{I. INTRODUCTION\bigskip}

Recently an exact proof is presented that the standard transformations (ST)
[1,2] (see also the standard textbooks, e.g. [3,4]) of the three-dimensional
(3D) vectors of the electric and magnetic fields, $\mathbf{E}$ and
$\mathbf{B}$ respectively, are not relativistically correct. This proof is
given both in the tensor formalism [5] and the geometric (Clifford) algebra
formalism [6]. It is shown in both formalisms that these ST of $\mathbf{E}$
and $\mathbf{B}$ drastically differ from the correct Lorentz transformations
(LT) of the corresponding 4D algebric objects representing the electric and
magnetic fields. The fundamental difference is that in the ST, e.g., the
components of the transformed 3D $\mathbf{E}^{\prime}$ are expressed by the
mixture of components of the 3D $\mathbf{E}$ and $\mathbf{B}$, and similarly
for $\mathbf{B}^{\prime}$. However, the correct LT always transform, e.g., the
electric field (i.e., the 4D algebric object representing the electric field)
only to the electric field, and similarly for the magnetic field. The
mentioned proof from [5,6] implies that the usual Maxwell equations (ME) with
the 3D $\mathbf{E}$ and $\mathbf{B}$ are not Lorentz covariant equations.
Consequently they are not equivalent to the field equations with the
electromagnetic field tensor $F^{ab}$ (tensor formalism) or to those with the
bivector field $F$ (the geometric algebra formalism). In this paper the above
statement will be exactly proved both in the geometric algebra and tensor
formalisms. Different 4D algebric objects are used to represent the standard
observer dependent and the new observer independent electric and magnetic
fields. First the electric and magnetic fields are represented by the
\emph{observer dependent }1-vectors $E_{f}$ and $B_{f}$ defined in the
$\gamma_{0}$ - frame. The usual ME in the component form are derived in
section II.A. and their LT are considered in section II.B. It is explicitly
shown in II.B., using the correct LT of $E_{f}$\ and $B_{f}$, that \emph{the
Lorentz transformed ME} \emph{are not of the same form as the original ones.}
This proves that, contrary to the general opinion, \emph{the usual ME are not
Lorentz covariant equations. }In section II.C. the ST of the usual ME are
considered taking into account the ST of the components of the 3D $\mathbf{E}$
and $\mathbf{B}$. It is proved that both the ST of the usual ME and the ST of
the 3D $\mathbf{E}$ and $\mathbf{B}$ have nothing in common with the correct
LT. The new Lorentz\ invariant\ field\ equations\ are constructed in section
II.D. in which the electric and magnetic fields are represented by the
\emph{observer independent}, i.e., defined without reference frames, 1-vectors
$E$\ and $B$. In sections III. to III.D. the whole consideration is repeated
but dealing with the \emph{observer dependent} bivectors $\mathbf{E}_{H}$ and
$\mathbf{B}_{H}$ defined in the $\gamma_{0}$ - frame and the \emph{observer
independent} bivectors $E_{HL}$ and $B_{HL}$.\ In the geometric algebra
formalism the active LT are used. Comparing the derivations in sections II. to
II.D. and sections III. to III.D. one concludes that the formulation with
1-vectors is simpler than the approach with bivectors and also it is much
closer to the classical formulation of the electromagnetism with the 3D
vectors $\mathbf{E}$ and $\mathbf{B}.$ In sections IV. to IV.D. the proof is
presented in the tensor formalism using the \emph{observer dependent}
4-vectors $E_{f}^{a}$ and $B_{f}^{a}$ defined in the $\gamma_{0}$ - frame and
the \emph{observer independent} 4-vectors $E^{a}$\textbf{\ }and\textbf{\ }%
$B^{a}$. In the tensor formalism the passive LT are used. All quantities in
the Lorentz invariant field equations derived with the use of 1-vectors
$E$\ and $B$, bivectors $E_{HL}$ and $B_{HL}$\ and the abstract 4-vectors
$E^{a}$\textbf{\ }and\textbf{\ }$B^{a}$ are geometric, coordinate-free
quantities, i.e., quantities that are defined without reference frames. All
such equations are completely equivalent to the field equations with $F$
(given, e.g. in [7-9] and discussed in detail in [10]) or with $F^{ab}$
(already presented, e.g., in [11]). It can be concluded from the consideration
presented in all mentioned sections that the proof of a fundamental
disagreement between the standard electromagnetism and the special relativity
(SR) does not depend on the character of the 4D algebric objects used to
represent the electric and magnetic fields. The discussion and a short
comparison with some experiments are given in section V. (We note that the
comparison of the geometric approach to SR and the standard formulation of SR
with experiments that test SR is also given in detail in [12].) The summary
and conclusions are presented in section VI. \bigskip\medskip

\textbf{II. THE PROOF IN THE GEOMETRIC\ ALGEBRA FORMALISM}

\textbf{USING\ 1-VECTORS\ }$E$\textbf{\ AND\ }$B$\bigskip

For the standard formulation of electrodynamics with the Clifford
multivectors, see, e.g., [7-9]. (A modern and very stimulating mathematical
treatment of the Clifford algebra and the geometric calculus is presented in
[13]. ) In [7-9] the electromagnetic field is represented by a bivector-valued
function $F=F(x)$ on the spacetime. The source of the field is the
electromagnetic current $j$ which is a 1-vector field and the gradient
operator $\partial$ is also 1-vector. A single field equation for $F$ is first
given by M. Riesz [14] as
\begin{equation}
\partial F=j/\varepsilon_{0}c,\quad\partial\cdot F+\partial\wedge
F=j/\varepsilon_{0}c. \label{MEF}%
\end{equation}
The trivector part is identically zero in the absence of magnetic charge. The
geometric (Clifford) product is written by simply juxtaposing multivectors
$AB$. The dot $^{\prime}\cdot^{\prime}$ and wedge $^{\prime}\wedge^{\prime}$
in (\ref{MEF}) denote the inner and outer products respectively. All
quantities in (\ref{MEF}) are defined without reference frames; they are
\emph{observer independent quantities}, i.e., they are independent of the
reference frame and the chosen system of coordinates in that frame.
Consequently the equation (\ref{MEF}) is a Lorentz invariant equation. In
fact, it is independent of even an indirect reference to an inertial system.
In the geometric algebra formalism (as in the tensor formalism as well) one
mainly deals either with 4D quantities that are defined without reference
frames, e.g., Clifford multivector $F$ (the abstract tensor $F^{ab}$) or, when
some basis has been introduced, with coordinate-based geometric quantity
(CBGQ) that comprises both components and a basis. The SR that exclusively
deals with quantities defined without reference frames or, equivalently, with
CBGQs, can be called the invariant SR. The reason for this name is that upon
the passive LT any CBGQ remains unchanged. The invariance of some 4D CBGQ upon
the passive LT reflects the fact that such mathematical, invariant, geometric
4D quantity represents \emph{the same physical object} for relatively moving
observers. \emph{It is taken in the invariant SR that such 4D geometric
quantities are well-defined not only mathematically but also experimentally,
as measurable quantities with real physical meaning. Thus they do have an
independent physical reality. }The invariant SR is discussed in [11,12] in the
tensor formalism and in [10,15] in the Clifford algebra formalism. It is
explicitly shown in [12] that the \emph{true agreement} with experiments that
test SR exists when the theory deals with well-defined 4D quantities, i.e.,
the quantities that are invariant upon the passive LT. The generally accepted
agreement between these experiments and the standard formulation of SR is only
an ''apparent'' agreement caused by the fact that in the standard treatments
only parts of the relevant 4D quantities are considered and thus not the whole
4D quantities, see [12].

In the usual geometric algebra formalism, e.g., [7-9], instead of working only
with such \emph{observer independent quantities} one introduces (in order to
get a more familiar form for (\ref{MEF})) a space-time split and the relative
vectors in the $\gamma_{0}$ - frame, i.e., a particular time-like direction
$\gamma_{0}$ is singled out. $\gamma_{0}$ is tangent to the world line of an
observer at rest in the $\gamma_{0}$ - frame.

(The generators of the spacetime algebra are four basis vectors $\left\{
\gamma_{\mu}\right\}  ,\mu=0...3,$ satisfying $\gamma_{\mu}\cdot\gamma_{\nu
}=\eta_{\mu\nu}=diag(+---).$ This basis is a right-handed orthonormal frame of
vectors in the Minkowski spacetime $M^{4}$ with $\gamma_{0}$ in the forward
light cone. The $\gamma_{k}$ ($k=1,2,3$) are spacelike vectors. The
$\gamma_{\mu}$ generate by multiplication a complete basis, the standard
basis, for spacetime algebra: $1,\gamma_{\mu},\gamma_{\mu}\wedge\gamma_{\nu
},\gamma_{\mu}\gamma_{5,}\gamma_{5}$ (16 independent elements). $\gamma_{5}$
is the pseudoscalar for the frame $\left\{  \gamma_{\mu}\right\}  .$ It is
worth noting that the standard basis corresponds, in fact, to the specific
system of coordinates, i.e., to Einstein's system of coordinates. In the
Einstein system of coordinates the Einstein synchronization [2] of distant
clocks and Cartesian space coordinates $x^{i}$ are used in the chosen inertial
frame of reference. However \emph{different systems of coordinates of an
inertial frame of reference are allowed and they are all equivalent in the
description of physical phenomena. }For example, in [11] two very different,
but completely equivalent systems of coordinates, the Einstein system of
coordinates and ''radio'' (''r'') system of coordinates, are exposed and
exploited throughout the paper. The connection between the basis vectors in
the ''r'' and in the Einstein system of coordinates is given as $r_{0}%
=\gamma_{0},\;r_{i}=\gamma_{0}+\gamma_{i}$. Thence the metric tensor
$g_{\mu\nu,r}$ in the ''r'' system of coordinates is given as $g_{00,r}%
=g_{0i,r}=g_{i0,r}=g_{ij,r}(i\neq j)=-1,\quad g_{ii,r}=0$; the metric tensor
$g_{\mu\nu,r}$ is not the same as the Minkowski metric tensor $\eta_{\mu\nu
}=diag(+---).$ We note that in SR, i.e., in the theory of flat spacetime, any
specific $g_{\mu\nu}$ (for the specific system of coordinates) can be
transformed to the Minkowski metric tensor $\eta_{\mu\nu}$; for example,
$g_{\mu\nu,r}$ is transformed by the matrix $(T_{\;\nu,r}^{\mu})^{-1}$ given
in [11] to $\eta_{\mu\nu}$. The coordinate system in which $g_{0i}=0$ at every
point in 4D spacetime is called time-orthogonal since in it the time axis is
everywhere orthogonal to the spatial coordinate curves. This happens in the
cases when in some inertial frame of reference the Einstein synchronization is
chosen together with, e.g., Cartesian, or polar, or spherical, etc., spatial
coordinates. However it is not the case when the ''r'' synchronization is
chosen. It is almost always tacitly assumed in both geometric algebra and
tensor formalisms that, e.g., for the spacetime algebra, [7] Space-Time
Calculus: ''a given inertial system is completely characterized by a single
future-pointing, timelike unit vector.'' In this case it refers to the unit
vector in the time direction, $\gamma_{0}$ basis vector, and the inertial
system characterized by $\gamma_{0}$ is refered to as the $\gamma_{0}$ -
frame, or the $\gamma_{0}$ - system. The preceding discussion shows that the
above claim from [7] is not true in general. Namely $\gamma_{0}$ and $r_{0}$
are the same vectors ($\gamma_{0}$, i.e., $r_{0},$ is the unit vector directed
along the world line of the clock at the origin), but the spatial basis
vectors $\gamma_{i}$ and $r_{i}$ are very different and moreover $r_{0}$ is
not orthogonal to $r_{i}$. (\emph{The spatial basis vectors }by definition
connect \emph{simultaneous} events, the event ''clock at rest at the origin
reads 0 time'' with the event ''clock at rest at unit distance from the origin
reads 0 time,'' and thus they \emph{are synchronization-dependent}. The
spatial basis vector $e_{i}$ connects two above mentioned simultaneous events
when Einstein's synchronization of distant clocks is used. The spatial basis
vector $r_{i}$ connects two above mentioned simultaneous events when ''radio''
clock synchronization of distant clocks is used. All this is explained in more
detail in [11].) This means that \emph{the usual space-time split and the
relative vectors, e.g., }[7-9]\emph{, are obtained not only by singling out a
particular time-like direction }$\gamma_{0}$ \emph{but also implicitly
assuming that the whole standard basis }$\left\{  \gamma_{\mu}\right\}
$\emph{\ (i.e., the Einstein system of coordinates)} \emph{is chosen}. In this
paper, for the sake of brevity and of clearness of the whole exposition, we
shall also work only with standard basis $\left\{  \gamma_{\mu}\right\}  $,
but remembering that the approach with 4D quantities that are defined without
reference frames holds for any choice of the basis.)

The bivector field $F$ is decomposed in the $\gamma_{0}$ - frame into electric
and magnetic parts using different algebric objects to represent these fields.
The explicit appearance of $\gamma_{0}$ in these expressions implies that
\emph{the space-time split is observer dependent} and thus all quantities
obtained by the space-time split in the $\gamma_{0}$ - frame are
\emph{observer dependent quantities}. In [7,8] the \emph{observer independent}
$F$ field from (\ref{MEF}) is expressed in terms of \emph{observer dependent
quantities,} i.e., as the sum of a relative vector $\mathbf{E}_{H}$ and a
relative bivector $\gamma_{5}\mathbf{B}_{H}$
\begin{align}
F  &  =\mathbf{E}_{H}+c\gamma_{5}\mathbf{B}_{H}\mathbf{,\quad E}_{H}%
=(F\cdot\gamma_{0})\gamma_{0}=(1/2)(F-\gamma_{0}F\gamma_{0}),\nonumber\\
\gamma_{5}\mathbf{B}_{H}  &  =(1/c)(F\wedge\gamma_{0})\gamma_{0}%
=(1/2c)(F+\gamma_{0}F\gamma_{0}). \label{FB}%
\end{align}
(The subscript 'H' is for - Hestenes.) Both $\mathbf{E}_{H}$ and
$\mathbf{B}_{H}$ are, in fact, bivectors. Similarly in [9] $F$ is decomposed
in terms of \emph{observer dependent quantities}, 1-vector $\mathbf{E}_{J}$
and a bivector $\mathbf{B}_{J}$ (the subscript 'J' is for - Jancewicz) as
$F=\gamma_{0}\wedge\mathbf{E}_{J}-c\mathbf{B}_{J},$ where $\mathbf{E}%
_{J}=F\cdot\gamma_{0}$ and $\mathbf{B}_{J}=-(1/c)(F\wedge\gamma_{0})\gamma
_{0}.$ The $F$ field can be also decomposed in terms of another algebric
objects; the \emph{observer dependent }electric and magnetic parts of $F$ are
represented with 1-vectors that are denoted as $E_{f}$ and $B_{f}$ (see also
[6] and [15]). The physical description with 1-vectors $E_{f}$ and $B_{f}$ is
simpler but \emph{completely equivalent }to the description with the bivectors
$\mathbf{E}_{H},$ $\mathbf{B}_{H}$ [7,8] or with 1-vector $\mathbf{E}_{J}$ and
a bivector $\mathbf{B}_{J}$ [9]. Such decomposition of $F$ is not only simpler
but also much closer to the classical representation of the electric and
magnetic fields by the 3D vectors $\mathbf{E}$ and $\mathbf{B}$ than those
used in [7-9]. Thus
\begin{align}
F  &  =E_{f}\wedge\gamma_{0}+c(\gamma_{5}B_{f})\cdot\gamma_{0},\nonumber\\
E_{f}  &  =F\cdot\gamma_{0},\ B_{f}=-(1/c)\gamma_{5}(F\wedge\gamma_{0}).
\label{ebg}%
\end{align}

Having at our disposal different decompositions of $F$ into \emph{observer
dependent quantities} we proceed to present the proof that the classical
electromagnetism and the SR are not in agreement first using the decomposition
(\ref{ebg}) and then (\ref{FB}). (We shall not deal with the decomposition of
$F$ into $\mathbf{E}_{J}$ and $\mathbf{B}_{J}$ from [9] since both the
procedure and the results are completely the same as with (\ref{ebg}) and
(\ref{FB}).)\bigskip\medskip

\textbf{A. The field equations in the} $\gamma_{0}$ - \textbf{frame.}
\textbf{The Maxwell equations}\bigskip

When (\ref{ebg}) is introduced into the field equation for $F$ (\ref{MEF}) we
find
\begin{align}
\partial\lbrack(F\cdot\gamma_{0})\wedge\gamma_{0}+(F\wedge\gamma_{0}%
)\cdot\gamma_{0}]  &  =j/\varepsilon_{0}c\nonumber\\
\partial(E_{f}\wedge\gamma_{0}+c(\gamma_{5}B_{f})\cdot\gamma_{0})  &
=j/\varepsilon_{0}c. \label{eqfi}%
\end{align}
The equations (\ref{eqfi}) can be now written as coordinate-based geometric
equations in the standard basis $\left\{  \gamma_{\mu}\right\}  $ and the
second equation becomes
\begin{align}
\{\partial_{\alpha}[\delta_{\quad\mu\nu}^{\alpha\beta}E_{f}^{\mu}(\gamma
_{0})^{\nu}+c\varepsilon^{\alpha\beta\mu\nu}(\gamma_{0})_{\mu}B_{f,\nu
}]-(j^{\beta}/\varepsilon_{0})\}\gamma_{\beta}+  & \nonumber\\
\partial_{\alpha}[\delta_{\quad\mu\nu}^{\alpha\beta}(\gamma_{0})^{\mu}%
cB_{f}^{\nu}+\varepsilon^{\alpha\beta\mu\nu}(\gamma_{0})_{\mu}E_{f,\nu}%
]\gamma_{5}\gamma_{\beta}  &  =0, \label{cl}%
\end{align}
where $\gamma_{0}=(\gamma_{0})^{\mu}\gamma_{\mu}$ with $(\gamma_{0})^{\mu
}=(1,0,0,0)$ and
\begin{align}
E_{f}  &  =E_{f}^{\mu}\gamma_{\mu}=0\gamma_{0}+F^{i0}\gamma_{i},\nonumber\\
B_{f}  &  =B_{f}^{\mu}\gamma_{\mu}=0\gamma_{0}+(-1/2c)\varepsilon^{0kli}%
F_{kl}\gamma_{i}. \label{gnl}%
\end{align}
Thence the components of $E_{f}$ and $B_{f}$ in the $\left\{  \gamma_{\mu
}\right\}  $ basis (i.e., in the Einstein system of coordinates) are
\begin{equation}
E_{f}^{i}=F^{i0},\quad B_{f}^{i}=(-1/2c)\varepsilon^{0kli}F_{kl}. \label{sko}%
\end{equation}
The relation (\ref{sko}) is nothing else than the standard identification of
the components $F^{\mu\nu}$ with the components of the 3D vectors $\mathbf{E}$
and $\mathbf{B,}$ see, e.g., [3,4]. (It is worth noting that Einstein's
fundamental work [16] is the earliest reference on covariant electrodynamics
and on the identification of some components of $F^{\alpha\beta}$ with the
components of the 3D $\mathbf{E}$ and $\mathbf{B.}$) We see that if the
standard basis $\left\{  \gamma_{\mu}\right\}  $ is chosen in an inertial
frame of reference, the $\gamma_{0}$ - frame, in which the observers who
measure the basis components $E_{f}^{\alpha}$ and $B_{f}^{\alpha}$ are at
rest, i.e., their velocity $v$ is $v=c\gamma_{0}$, or in the components
$v^{\alpha}=(c,0,0,0)$, then in the $\gamma_{0}$ - frame $E_{f}$ \emph{and}
$B_{f}$ \emph{do not have the temporal components} $E_{f}^{0}=B_{f}^{0}=0$.
Thus $E_{f}$ and $B_{f}$ actually refer to the 3D subspace orthogonal to the
specific timelike direction $\gamma_{0}.$ Notice that we can select a
particular - but otherwise arbitrary - inertial frame of reference as the
$\gamma_{0}$ - frame, to which we shall refer as the frame of our 'fiducial'
observers (for this name see [17]). The subscript $^{\prime}f^{\prime}$ in the
above relations stands for - fiducial - and denotes the explicit dependence of
these quantities on the $\gamma_{0}$-, i.e., 'fiducial'- observer. Using that
$E_{f}^{0}=B_{f}^{0}=0$ and $(\gamma_{0})^{\mu}=(1,0,0,0)$ the equation
(\ref{cl}) becomes
\begin{align}
(\partial_{k}E_{f}^{k}-j^{0}/c\varepsilon_{0})\gamma_{0}+(-\partial_{0}%
E_{f}^{i}+c\varepsilon^{ijk0}\partial_{j}B_{fk}-j^{i}/c\varepsilon_{0}%
)\gamma_{i}+  & \nonumber\\
(-c\partial_{k}B_{f}^{k})\gamma_{5}\gamma_{0}+(c\partial_{0}B_{f}%
^{i}+\varepsilon^{ijk0}\partial_{j}E_{fk})\gamma_{5}\gamma_{i}  &  =0.
\label{MEC}%
\end{align}
The first part (with $\gamma_{\alpha}$) in (\ref{MEC}) is from the 1-vector
part of (\ref{eqfi}), i.e., (\ref{cl}), while the second one (with $\gamma
_{5}\gamma_{\alpha}$) is from the trivector (pseudovector) part of
(\ref{eqfi}), i.e., (\ref{cl}). Both parts in (\ref{MEC}) are written as
coordinate-based geometric equations in the standard basis $\left\{
\gamma_{\mu}\right\}  $ and cannot be further simplified as geometric
equations. In the first part (with $\gamma_{\alpha}$) in (\ref{MEC}) one
recognizes \emph{two} \emph{Maxwell equations} in the \emph{component form},
the Gauss law for the electric field (the first bracket, with $\gamma_{0}$)
and the Amp\`{e}re-Maxwell law (the second bracket, with $\gamma_{i}$).
Similarly from the second part (with $\gamma_{5}\gamma_{\alpha}$) in
(\ref{MEC}) we recognize the \emph{component form} of another \emph{two
Maxwell equations}, the Gauss law for the magnetic field (with $\gamma
_{5}\gamma_{0}$) and Faraday's law (with $\gamma_{5}\gamma_{i}$).\medskip\bigskip

\textbf{B. Lorentz transformations of the Maxwell equations} \bigskip

Let us now apply the active Lorentz transformations upon (\ref{MEC}), or
(\ref{cl}). We write (\ref{MEC}), or (\ref{cl}), in the form
\begin{equation}
a^{\alpha}\gamma_{\alpha}+b^{\alpha}(\gamma_{5}\gamma_{\alpha})=0. \label{ab}%
\end{equation}
The coefficients $a^{\alpha}$ and $b^{\alpha}$ are clear from (\ref{MEC}), or
(\ref{cl}); they are the usual Maxwell equations in the component form. In the
Clifford algebra formalism, e.g., [7-9], the LT are considered as active
transformations; the components of, e.g., some 1-vector relative to a given
inertial frame of reference (with the standard basis $\left\{  \gamma_{\mu
}\right\}  $) are transformed into the components of a new 1-vector relative
to the same frame (the basis $\left\{  \gamma_{\mu}\right\}  $ is not
changed). Furthermore the LT are described with rotors $R,$ $R\widetilde
{R}=1,$ in the usual way as $p\rightarrow p^{\prime}=Rp\widetilde{R}%
=p^{\prime\mu}\gamma_{\mu}.$ To an observer in the $\left\{  \gamma_{\mu
}\right\}  $ frame the vector $p^{\prime}$ appears the same as the vector $p$
appears to an observer in the $\left\{  \gamma_{\mu}^{\prime}\right\}  $
frame. For boosts in the direction $\gamma_{1}$ the rotor $R$ is given by the
relation
\begin{equation}
R=(1+\gamma+\gamma\beta\gamma_{0}\gamma_{1})/(2(1+\gamma))^{1/2}, \label{err}%
\end{equation}
$\beta$ is the scalar velocity in units of $c$, $\gamma=(1-\beta^{2})^{-1/2}.$
Then the LT of (\ref{eqfi}) are given as
\begin{align}
R\{\partial\lbrack(F\cdot\gamma_{0})\wedge\gamma_{0}+(F\wedge\gamma_{0}%
)\cdot\gamma_{0}]-j/\varepsilon_{0}c\}\widetilde{R}  &  =0,\nonumber\\
R\{\partial\lbrack E_{f}\wedge\gamma_{0}+c(\gamma_{5}B_{f})\cdot\gamma
_{0}]-j/\varepsilon_{0}c\}\widetilde{R}  &  =0, \label{rem}%
\end{align}
where $R$ is given by (\ref{err}). (A coordinate-free form of the LT is also
given in the Clifford algebra formalism in [15] and in the tensor formalism in
[11]. The form presented in [15] does not need to use rotors but, of course,
it can be expressed by rotors as well.) Then the LT of (\ref{ab}) are
\begin{equation}
R\{a^{\alpha}\gamma_{\alpha}+b^{\alpha}(\gamma_{5}\gamma_{\alpha}%
)\}\widetilde{R}=0. \label{lab}%
\end{equation}
Performing the LT we find the explicit expression for (\ref{lab}) as
\begin{align}
\gamma_{0}(\gamma a^{0}-\beta\gamma a^{1})+\gamma_{1}(\gamma a^{1}-\beta\gamma
a^{0})+\gamma_{2}a^{2}+\gamma_{3}a^{3}+  & \nonumber\\
\gamma_{5}\gamma_{0}(\gamma b^{0}-\beta\gamma b^{1})+\gamma_{5}\gamma
_{1}(\gamma b^{1}-\beta\gamma b^{0})+\gamma_{5}\gamma_{2}b^{2}+\gamma
_{5}\gamma_{3}b^{3}  &  =0. \label{L}%
\end{align}
It can be simply written as
\begin{equation}
a^{\prime\alpha}\gamma_{\alpha}+b^{\prime\alpha}(\gamma_{5}\gamma_{\alpha})=0,
\label{L1}%
\end{equation}
where, e.g., $a^{\prime0}=\gamma a^{0}-\beta\gamma a^{1}$ and, as it is said,
$a^{\alpha}$ and $b^{\alpha}$ are the usual Maxwell equations in the component
form given in (\ref{MEC}), or (\ref{cl}). This result (\ref{L}), i.e.,
(\ref{L1}), is exactly the usual result for the active LT of a 1-vector and of
a pseudovector. It is important to note that, e.g., the Gauss law for the
electric field $a^{0}$ does not transform by the LT again to the Gauss law but
to $a^{\prime0}$, which is a combination of the Gauss law and a part of the
Amp\`{e}re-Maxwell law ($a^{1}$).

The second equation in (\ref{rem}) can be expressed in terms of Lorentz
transformed derivatives and Lorentz transformed 1-vectors $E_{f}$ and $B_{f}$
as
\begin{equation}
\partial^{\prime}[E_{f}^{\prime}\wedge\gamma_{0}^{\prime}+c(\gamma_{5}%
B_{f}^{\prime})\cdot\gamma_{0}^{\prime}]-j^{\prime}/\varepsilon_{0}c=0,
\label{mcr}%
\end{equation}
where $\partial^{\prime}=R\partial\widetilde{R},$ $\gamma_{0}^{\prime}%
=R\gamma_{0}\widetilde{R}=\gamma\gamma_{0}-\beta\gamma\gamma_{1}$ and (see
also [6]) the Lorentz transformed $E_{f}^{\prime}$ is
\begin{align}
E_{f}^{\prime}  &  =R(F\cdot\gamma_{0})\widetilde{R}=RE_{f}\widetilde
{R}=R(F^{i0}\gamma_{i})\widetilde{R}=E_{f}^{\prime\mu}\gamma_{\mu}=\nonumber\\
&  =-\beta\gamma E_{f}^{1}\gamma_{0}+\gamma E_{f}^{1}\gamma_{1}+E_{f}%
^{2}\gamma_{2}+E_{f}^{3}\gamma_{3}, \label{nle}%
\end{align}
what is the usual form for the active LT of the 1-vector $E_{f}=E_{f}^{\mu
}\gamma_{\mu}$. Similarly is obtained for $B_{f}^{\prime}$
\begin{align}
B_{f}^{\prime}  &  =R\left[  -(1/c)\gamma_{5}(F\wedge\gamma_{0})\right]
\widetilde{R}=RB_{f}\widetilde{R}=R\left[  (-1/2c)\varepsilon^{0kli}%
F_{kl}\gamma_{i}\right]  \widetilde{R}=\nonumber\\
&  =B_{f}^{\prime\mu}\gamma_{\mu}=-\beta\gamma B_{f}^{1}\gamma_{0}+\gamma
B_{f}^{1}\gamma_{1}+B_{f}^{2}\gamma_{2}+B_{f}^{3}\gamma_{3}. \label{nlb}%
\end{align}
It is worth noting that $E_{f}^{\prime}$ \emph{and} $B_{f}^{\prime}$ \emph{are
not more orthogonal to} $\gamma_{0},$ i.e., \emph{they} \emph{do have}
\emph{the temporal components} $\neq0.$ Furthermore \emph{the components}
$E_{f}^{\mu}$ ($B_{f}^{\mu}$) \emph{transform upon the active LT again to the
components} $E_{f}^{\prime\mu}$ ($B_{f}^{\prime\mu}$) as seen from (\ref{nle})
and (\ref{nlb}); \emph{there is no mixing of components}. When (\ref{mcr}) is
written in an expanded form as a coordinate-based geometric equation in the
standard basis $\left\{  \gamma_{\mu}\right\}  $ it takes the form of
(\ref{L1}) but now the coefficients $a^{\prime\alpha}$ are written by means of
the Lorentz transformed components $\partial_{k}^{\prime}$, $E_{f}^{\prime k}$
and $B_{f}^{\prime k}$ (for simplicity only the term $a^{\prime0}\gamma_{0}$
is presented)
\begin{equation}
a^{\prime0}\gamma_{0}=\{[\gamma(\partial_{k}^{\prime}E_{f}^{\prime
k})-j^{\prime0}/c\varepsilon_{0}]+\beta\gamma\lbrack\partial_{1}^{\prime}%
E_{f}^{\prime0}+c(\partial_{2}^{\prime}B_{f3}^{\prime}-\partial_{3}^{\prime
}B_{f2}^{\prime})]\}\gamma_{0}, \label{anu}%
\end{equation}
and \emph{it substantially differs in form from the term} $a^{0}\gamma
_{0}=(\partial_{k}E_{f}^{k}-j^{0}/c\varepsilon_{0})\gamma_{0}$ in (\ref{MEC}).
As explained above the coefficient $a^{0}$ is the Gauss law for the electric
field written in the component form. It is clear from (\ref{anu}) that the LT
do not transform the Gauss law into the 'primed' Gauss law but into quite
different law (\ref{anu}); $a^{\prime0}$ contains the time component
$E_{f}^{\prime0}$ (while $E_{f}^{0}=0$), and also the new ''Gauss law''
includes the derivatives of the magnetic field. The same situation happens
with other Lorentz transformed terms, which explicitly shows that \emph{the
Lorentz transformed ME} ((\ref{mcr}) \emph{with} (\ref{anu})) \emph{are not of
the same form as the original ones} (\ref{MEC}). This is a fundamental result
which reveals that, contrary to the previous derivations, e.g., [2,16], [3,4],
[7-9], and contrary to the general opinion, \emph{the usual ME are not Lorentz
covariant equations. }The physical consequences of this achievement will be
very important and they will be carefully examined.\bigskip\medskip

\textbf{C. Standard transformations of the Maxwell equations} \bigskip

In contrast to the correct active Lorentz transformations of $E_{f}$
(\ref{nle}) and $B_{f}$ (\ref{nlb}) it is wrongly assumed in the usual
derivations of the \emph{the ST for} $E_{st}^{\prime}$ \emph{and}
$B_{st}^{\prime}$ (the subscript - st - is for - standard)\emph{\ that the
quantities obtained by the active LT of} $E_{f}$ \emph{and} $B_{f}$ \emph{are
again in the 3D subspace of the} $\gamma_{0}$ \emph{-} \emph{observer }(see
also [6]). This means that it is wrongly assumed in all standard derivations,
e.g., in the Clifford algebra formalism [7,9] (and in the tensor formalism
[3,4] as well), that one can again perform the same identification of the
transformed components $F^{\prime\mu\nu}$ with the components of the 3D
$\mathbf{E}^{\prime}$ and $\mathbf{B}^{\prime}$ as in (\ref{sko}). Thus it is
taken in standard derivations that for the transformed $E_{st}^{\prime}$ and
$B_{st}^{\prime}$ again hold $E_{st}^{\prime0}=B_{st}^{\prime0}=0$ as for
$E_{f}$ and $B_{f}$,
\begin{align}
E_{st}^{\prime}  &  =(RF\widetilde{R})\cdot\gamma_{0}=F^{\prime}\cdot
\gamma_{0}=F^{\prime i0}\gamma_{i}=E_{st}^{\prime i}\gamma_{i}=\nonumber\\
&  =E_{f}^{1}\gamma_{1}+(\gamma E_{f}^{2}-\beta\gamma cB_{f}^{3})\gamma
_{2}+(\gamma E_{f}^{3}+\beta\gamma cB_{f}^{2})\gamma_{3}, \label{ce}%
\end{align}
where $F^{\prime}=RF\widetilde{R}$, and similarly for $B_{st}^{\prime}$
\begin{align}
B_{st}^{\prime}  &  =-(1/c)\gamma_{5}(F^{\prime}\wedge\gamma_{0}%
)=-(1/2c)\varepsilon^{0kli}F_{kl}^{\prime}\gamma_{i}=B_{st}^{\prime i}%
\gamma_{i}=\nonumber\\
&  B_{f}^{1}\gamma_{1}+(\gamma B_{f}^{2}+\beta\gamma E_{f}^{3}/c)\gamma
_{2}+(\gamma B_{f}^{3}-\beta\gamma E_{f}^{2}/c)\gamma_{3}. \label{B}%
\end{align}
From \emph{the relativistically incorrect transformations} (\ref{ce}) and
(\ref{B}) one simply finds the transformations of the spatial components
$E_{st}^{\prime i}$ and $B_{st}^{\prime i}$
\begin{equation}
E_{st}^{\prime i}=F^{\prime i0},\quad B_{st}^{\prime i}=(-1/2c)\varepsilon
^{0kli}F_{kl}^{\prime}. \label{sk1}%
\end{equation}
As can be seen from (\ref{ce}) and (\ref{B}), i.e., from (\ref{sk1}),
\emph{the transformations for} $E_{st.}^{\prime i}$ \emph{and} $B_{st.}%
^{\prime i}$ \emph{are exactly the ST of components of the 3D vectors}
$\mathbf{E}$ \emph{and} $\mathbf{B}$ that are quoted in almost every textbook
and paper on relativistic electrodynamics including [2] and [3,4]. These
relations are explicitly derived and given in the Clifford algebra formalism,
e.g., in [7], Space-Time Algebra (eq. (18.22)), New Foundations for Classical
Mechanics (Ch. 9 eqs. (3.51a,b)), in [8] Geometric algebra for physicists (Ch.
7.1.2 eq. (7.33)) and in [9] (Ch. 7 eqs. (20a,b)). Notice that, in contrast to
the active Lorentz transformations (\ref{nle}) and (\ref{nlb}),
\emph{according to the ST} (\ref{ce}), \emph{i.e.}, (\ref{sk1}), \emph{the
transformed components} $E_{st}^{\prime i}$ \emph{are expressed by the mixture
of components} $E_{f}^{i}$ \emph{and} $B_{f}^{i},$ \emph{and }(\ref{B}%
)\emph{\ shows that the same holds for} $B_{st}^{\prime i}$. In all previous
treatments of SR, e.g., [7-9] (and [2-4]) the transformations for
$E_{st.}^{\prime i}$ and $B_{st.}^{\prime i}$ are considered to be the Lorentz
transformations of the 3D electric and magnetic fields. However the above
analysis, and [5,6] as well, show that the transformations for $E_{st.}%
^{\prime i}$ and $B_{st.}^{\prime i}$ (\ref{sk1}) are derived from \emph{the
relativistically incorrect transformations} (\ref{ce}) and (\ref{B}), which
are not the Lorentz transformations; the Lorentz transformations are given by
the relations (\ref{nle}) and (\ref{nlb}).

It is also argued in all previous works, starting in the year 1905 with
Einstein's fundamental paper on SR [2], that the usual ME with the 3D
$\mathbf{E}$ and $\mathbf{B}$ are Lorentz covariant equations. The relation
(\ref{mcr}) together with (\ref{anu}) shows that it is not true; the Lorentz
transformed ME are not of the same form as the original ones. Here we
explicitly show that \emph{in the standard derivations} \emph{the ME remain
unchanged in form not upon the LT but upon some transformations which,
strictly speaking, have nothing to do with the LT of the equation}
(\ref{eqfi}), \emph{i.e., of the usual ME} (\ref{MEC}). The difference between
the Lorentz transformed ME, given by (\ref{rem}) or finally by (\ref{mcr})
with (\ref{anu}) (or by (\ref{L})) and the equations (given below) obtained by
applying the ST is the same as it is the difference between the LT of $E_{f}$
($B_{f}$) given by (\ref{nle}) ((\ref{nlb})) and their ST given by (\ref{ce})
((\ref{B})). Thus the ST of the equation (\ref{eqfi}) are
\begin{align}
(R\partial\widetilde{R})\{[(RF\widetilde{R})\cdot\gamma_{0}]\wedge\gamma
_{0}+[(RF\widetilde{R})\wedge\gamma_{0}]\cdot\gamma_{0}\}-(Rj\widetilde
{R})/\varepsilon_{0}c  &  =0,\nonumber\\
\partial^{\prime}\{E_{st}^{\prime}\wedge\gamma_{0}+c(\gamma_{5}B_{st}^{\prime
})\cdot\gamma_{0}\}-j^{\prime}/\varepsilon_{0}c  &  =0, \label{rtr}%
\end{align}
where $E_{st}^{\prime}$ and $B_{st}^{\prime}$ are defined by (\ref{ce}) and
(\ref{B}). Notice that, in contrast to the correct LT (\ref{rem}) or
(\ref{mcr}), $\gamma_{0}$ \emph{is not transformed in} (\ref{rtr}). The second
equation in (\ref{rtr}) is of the same form as the second equation in
(\ref{eqfi}) but with primed derivative $\partial^{\prime}$, $E_{st}^{\prime}$
and $B_{st}^{\prime}$ fields and the primed current $j^{\prime}$ replacing the
corresponding unprimed quantities. When this second equation in (\ref{rtr}) is
written as a coordinate-based geometric equation in the standard basis
$\left\{  \gamma_{\mu}\right\}  $ it becomes
\begin{align}
(\partial_{k}^{\prime}E_{st}^{\prime k}-j^{\prime0}/c\varepsilon_{0}%
)\gamma_{0}+(-\partial_{0}^{\prime}E_{st}^{\prime i}+c\varepsilon
^{ijk0}\partial_{j}^{\prime}B_{st,k}^{\prime}-j^{\prime i}/c\varepsilon
_{0})\gamma_{i}+  & \nonumber\\
(-c\partial_{k}^{\prime}B_{st}^{\prime k})\gamma_{5}\gamma_{0}+(c\partial
_{0}^{\prime}B_{st}^{\prime i}+\varepsilon^{ijk0}\partial_{j}^{\prime}%
E_{st,k}^{\prime})\gamma_{5}\gamma_{i}  &  =0. \label{EBC}%
\end{align}
The equation (\ref{EBC}) is of the same form as the original ME (\ref{MEC})
but \emph{the electric and magnetic fields are not transformed by the LT than
by the ST}. Therefore, as can be seen from (\ref{rtr}) (together with
(\ref{ce}) and (\ref{B})), \emph{the equation }(\ref{EBC})\emph{\ is not the
LT of the original ME} (\ref{MEC}); \emph{the LT of the ME} (\ref{MEC})
\emph{are the equations} (\ref{mcr}) \emph{with} (\ref{anu}) \emph{(i.e.,}
(\ref{L})) where the Lorentz transformed electric and magnetic fields are
given by the relations (\ref{nle}) and (\ref{nlb}). \bigskip\medskip

\textbf{D. Lorentz\ invariant\ field\ equations\ with 1-vectors }%
$E$\textbf{\ and }$B$ \bigskip

Instead of to decompose $F$ into the \emph{observer dependent }$E_{f}$ and
$B_{f}$ in the $\gamma_{0}$ - frame, as in (\ref{ebg}), we present here
\emph{an observer independent} decomposition of $F$ into 1-vectors of the
electric $E$ and magnetic $B$ fields that are defined without reference
frames, i.e., \emph{they are independent of the chosen reference frame and of
the chosen system of coordinates in it,} see also [15]$.$ We define
\begin{align}
F  &  =(1/c)E\wedge v+(e_{5}B)\cdot v,\nonumber\\
E  &  =(1/c)F\cdot v,\quad e_{5}B=(1/c^{2})F\wedge v,\ B=-(1/c^{2}%
)e_{5}(F\wedge v), \label{myF}%
\end{align}
where the pseudoscalar $e_{5}$ of some basis $\left\{  e_{\mu}\right\}  $,
that does not need to be the standard basis $\left\{  \gamma_{\mu}\right\}  $,
is defined as $e_{5}=e_{0}\wedge e_{1}\wedge\wedge e_{2}\wedge e_{3}.$ It
holds that $E\cdot v=B\cdot v=0$ (since $F$ is skew-symmetric). $v$ in
(\ref{myF}) can be interpreted as the velocity (1-vector) of a family of
observers who measures $E$ and $B$ fields. \emph{The velocity} $v$ \emph{and
all other quantities entering into} (\ref{myF}) \emph{are defined without
reference frames.} $v$ characterizes some general observer. Thus \emph{both
relations }in (\ref{myF})\emph{\ hold for any observer.} However it has to be
emphasized that (\ref{myF}) is not a physical definition of $E$ and $B;$ the
physical definition has to be given in terms of the Lorentz force and Newton's
second law as, e.g., in [15]. The relations (\ref{myF}) actually establish the
equivalence of the formulation of electrodynamics with the field bivector $F$
and the formulation with 1-vectors of the electric $E$ and magnetic $B$
fields. (Recently [10] I have presented a complete formulation of the
electrodynamics using exclusively the bivector field $F.$) \emph{Both
formulations, with} $F$ \emph{and} $E,$ $B$ \emph{fields, are equivalent
formulations, but every of them is a complete, consistent and self-contained
formulation.} When (\ref{myF}) is used the field equation for $F$ (\ref{MEF})
becomes
\begin{equation}
\partial((1/c)E\wedge v+(e_{5}B)\cdot v)=j/\varepsilon_{0}c. \label{deb}%
\end{equation}
In contrast to the field equation (\ref{eqfi}), that holds only for the
$\gamma_{0}$-observer, the field equation (\ref{deb}) \emph{holds for any
observer;} \emph{the quantities entering into} (\ref{deb}) \emph{are all
defined without reference frames.} \emph{The equation} (\ref{deb}) \emph{is
physicaly completely equivalent to the field equation for} $F$ (\ref{MEF}). In
some basis $\left\{  e_{\mu}\right\}  ,$ that does not need to be the standard
basis $\left\{  \gamma_{\mu}\right\}  ,$ the field equation (\ref{deb}) can be
written as a coordinate-based geometric equation
\begin{align}
\lbrack\partial_{\alpha}(\delta_{\quad\mu\nu}^{\alpha\beta}E^{\mu}v^{\nu
}+\varepsilon^{\alpha\beta\mu\nu}v_{\mu}cB_{\nu})-  &  (j^{\beta}%
/\varepsilon_{0})]e_{\beta}+\nonumber\\
\partial_{\alpha}(\delta_{\quad\mu\nu}^{\alpha\beta}v^{\mu}cB^{\nu
}+\varepsilon^{\alpha\beta\mu\nu}v_{\mu}E_{\nu})e_{5}e_{\beta}  &  =0,
\label{maeb}%
\end{align}
where $E^{\alpha}$ and $B^{\alpha}$ are the basis components of the electric
and magnetic 1-vectors $E$ and $B$, and $\delta_{\quad\mu\nu}^{\alpha\beta
}=\delta_{\,\,\mu}^{\alpha}\delta_{\,\,\nu}^{\beta}-\delta_{\,\,\nu}^{\alpha
}\delta_{\,\mu}^{\beta}.$ The first part in (\ref{maeb}) (it contains sources)
emerges from $\partial\cdot F=j/\varepsilon_{0}c$ and the second one (the
source-free part) is obtained from $\partial\wedge F=0,$ see also [15]$.$
Instead of to work with the observer independent field equation in the $F$-
formulation (\ref{MEF}) one can equivalently use the $E,$ $B$ - formulation
with the field equation (\ref{deb}), or in the $\left\{  e_{\mu}\right\}  $
basis (\ref{maeb}). (The complete $E,$ $B$ - formulation of the relativistic
electrodynamics will be reported elsewhere.) We remark that (\ref{maeb})
follows from (\ref{deb}) for those systems of coordinates for which the basis
1-vectors $e_{\mu}$ are constant, e.g., the standard basis $\left\{
\gamma_{\mu}\right\}  $ (the Einstein system of coordinates). For a
nonconstant basis, for example, when one uses polar or spherical basis 1-
vectors (and, e.g., the Einstein synchronization) then one must also
differentiate these nonconstant basis 1-vectors. Furthermore one can
completely forget the manner in which the equation with $E$ and $B$ is
obtained, i.e., the field equation with $F$ (\ref{MEF}), \emph{and consider
the equation with} $E$ \emph{and} $B$ (\ref{deb}), \emph{which is defined
without reference frames, or the corresponding coordinate-based geometric
equation} (\ref{maeb}), \emph{as the primary and fundamental equations for the
whole classical electromagnetism. }In such correct relativistic formulation of
the electromagnetism the field equation with 1- vectors $E$ and $B$
(\ref{deb}) takes over the role of the usual ME with the 3D $\mathbf{E}$ and
$\mathbf{B}$, i.e., of the ME (\ref{MEC}). We note that the equivalent
formulation of electrodynamics with tensors $E^{a}$ and $B^{a}$ is reported in
[11,18] while the component form in the Einstein system of coordinates\ is
given in [17,19] and [20].

Let us now take that in (\ref{maeb}) the standard basis $\left\{  \gamma_{\mu
}\right\}  $ is used instead of some general basis $\left\{  e_{\mu}\right\}
.$ Then (\ref{maeb}) can be written as $C^{\beta}\gamma_{\beta}+D^{\beta
}\gamma_{5}\gamma_{\beta}=0,$ where $C^{\beta}=\partial_{\alpha}(\delta
_{\quad\nu\mu}^{\alpha\beta}v^{\mu}E^{\nu}-\varepsilon^{\alpha\beta\nu\mu
}v_{\mu}cB_{\nu})-j^{\beta}/\varepsilon_{0}$ and $D^{\beta}=\partial_{\alpha
}(\delta_{\quad\mu\nu}^{\alpha\beta}v^{\mu}cB^{\nu}+\varepsilon^{\alpha
\beta\mu\nu}v_{\mu}E_{\nu})$. \emph{When the active LT are applied upon such}
(\ref{maeb}) \emph{with the} $\left\{  \gamma_{\mu}\right\}  $ \emph{basis the
equation remains of the same form but with primed quantities replacing the
unprimed ones (of course the basis is unchanged).} This can be immediately
seen since the equation (\ref{maeb}) is written in a manifestly covariant
form. Thus the Lorentz transformed (\ref{maeb}) is
\begin{align}
R(C^{\beta}\gamma_{\beta}+D^{\beta}\gamma_{5}\gamma_{\beta})\widetilde{R}  &
=0,\nonumber\\
C^{\prime\beta}\gamma_{\beta}+D^{\prime\beta}\gamma_{5}\gamma_{\beta}  &  =0,
\label{clo}%
\end{align}
where, e.g., $C^{\prime\beta}=\partial_{\alpha}^{\prime}(\delta_{\quad\nu\mu
}^{\alpha\beta}v^{\prime\mu}E^{\prime\nu}-\varepsilon^{\alpha\beta\nu\mu
}v_{\mu}^{\prime}cB_{\nu}^{\prime})-j^{\prime\beta}/\varepsilon_{0}$.
Obviously \emph{such formulation of the electromagnetism with fundamental
equation} (\ref{deb}) \emph{or} (\ref{maeb}) \emph{is a relativistically
correct formulation.}

What is the relation between the relativistically correct field equation
(\ref{deb}) or (\ref{maeb}) and the usual ME (\ref{MEC}). From the above
discussion and from section II.A. one concludes that if in (\ref{deb}) we
specify the velocity $v$ of the observers who measure $E$ and $B$\ fields to
be $v=c\gamma_{0}$, then the equation (\ref{deb}) becomes the equation
(\ref{eqfi}). Further choosing the standard basis $\left\{  \gamma_{\mu
}\right\}  $ in the $\gamma_{0}$ - frame, in which $v=c\gamma_{0}$, or in the
components $v^{\alpha}=(c,0,0,0)$, then in that $\gamma_{0}$ - frame $E$ and
$B$ become $E_{f}$ and $B_{f}$ and they do not have the temporal components
$E_{f}^{0}=B_{f}^{0}=0$. \emph{The coordinate-based geometric equation}
(\ref{maeb}) \emph{becomes the usual Maxwell equations} (\ref{MEC}). Thus the
usual Clifford algebra treatments of the electromagnetism [7-9] with the
space-time split in the $\gamma_{0}$ - frame and the usual ME (\ref{MEC}) are
simply obtained from our \emph{observer independent} formulation with field
equation (\ref{deb}) or (\ref{maeb}) choosing that $v=c\gamma_{0} $ and
choosing the standard basis $\left\{  \gamma_{\mu}\right\}  $. We see that the
correspondence principle is simply satisfied in this formulation with $E$ and
$B$\ fields; all results obtained in the previous treatments from the usual
Maxwell equations with the 3D $\mathbf{E}$ and $\mathbf{B}$ remain valid in
the formulation with the 1-vectors $E$ and $B$ if physical phenomena are
considered only in one inertial frame of reference. Namely the selected
inertial frame of reference can be chosen to be the $\gamma_{0}$ - frame with
the $\left\{  \gamma_{\mu}\right\}  $ basis. Then there, as explained above,
the coordinate-based geometric equation (\ref{maeb}) can be reduced to the
equations containing only the components, the four Maxwell equations in the
component form, the ME (\ref{MEC}). Thus for observers who are at rest in the
$\gamma_{0}$ - frame ($v=c\gamma_{0}$) the components of the 3D $\mathbf{E}$
and $\mathbf{B}$ can be simply replaced by the space components of the
1-vectors $E$ and $B$ in the $\left\{  \gamma_{\mu}\right\}  $ basis. We
remark that just such observers are usually considered in the conventional
formulation with the 3D $\mathbf{E}$ and $\mathbf{B.}$ The dependence of the
field equations (\ref{maeb}) on $v$ reflects the arbitrariness in the
selection of the $\gamma_{0}$ - frame but at the same time it makes the
equations (\ref{maeb}) independent of that choice. The $\gamma_{0}$ - frame
can be selected at our disposal, which proves that we don't have a kind of the
''preferred'' frame theory. \emph{All experimental results that are obtained
in one inertial frame of reference can be equally well explained by our
geometric formulation of electrodynamics with the 1-vectors} $E$ \emph{and}
$B$ \emph{as they are explained by the usual ME with the 3D} $\mathbf{E}$
\emph{and }$\mathbf{B.}$

However there is a fundamental difference between the standard approach with
the 3D $\mathbf{E}$ and $\mathbf{B}$ and the approach with 4D quantities $E$
and $B$ that are defined without reference frames. It is considered in all
standard treatments that the equation (\ref{EBC}) is the LT\ of the original
ME (\ref{MEC}). But, as shown here, the equation (\ref{EBC}) is not the LT of
the original ME (\ref{MEC}); the LT of the ME (\ref{MEC}) are the equations
(\ref{L}) (i.e., (\ref{L1}) with (\ref{anu}), or (\ref{mcr})). The ME
(\ref{MEC}) are obtained from our field equation (\ref{maeb}) by putting that
$v=c\gamma_{0}$ and choosing the standard basis $\left\{  \gamma_{\mu
}\right\}  $. In the same way the equations (\ref{clo}), which are the LT of
the equations (\ref{maeb}), become the LT of the ME (\ref{MEC}), that is, the
equations (\ref{L}) (or (\ref{L1}) with (\ref{anu}), or (\ref{mcr})), when in
(\ref{clo}) it is taken that $v^{\prime}$, $\partial^{\prime}$, $E^{\prime}$
and $B^{\prime}$ are the LT of $v=c\gamma_{0}$, $\partial$, $E_{f}$ and
$B_{f}$, that is, $v^{\prime}=R(c\gamma_{0})\widetilde{R}$, $\partial^{\prime
}=R\partial\widetilde{R}$, $E^{\prime}=RE_{f}\widetilde{R}=E_{f}^{\prime},$
$B^{\prime}=RB_{f}\widetilde{R}$=$B_{f}^{\prime}$. We recall from section
II.B. that to an observer in the $\left\{  \gamma_{\mu}\right\}  $ frame the
vector $p^{\prime}$ ($p^{\prime}=Rp\widetilde{R}=p^{\prime\mu}\gamma_{\mu}$)
appears the same as the vector $p$ ($p=p^{\mu}\gamma_{\mu}$) appears to an
observer in the $\left\{  \gamma_{\mu}^{\prime}\right\}  $ frame. This,
together with the preceding discussion, show that the usual ME with the 3D
$\mathbf{E}$ and $\mathbf{B}$, i.e., the equation (\ref{MEC}) and the equation
(\ref{EBC}) obtained by the ST from (\ref{MEC}), cannot be used for the
explanation of any experiment that test SR, i.e., in which relatively moving
observers have to compare their data \emph{obtained by measurements on the
same physical object.} In contrast to the description of the electromagnetism
with the 3D $\mathbf{E}$ and $\mathbf{B,}$ \emph{the description with 4D
fields }$E$ \emph{and} $B$, \emph{i.e., with the equations }(\ref{maeb})
\emph{and} (\ref{clo}), \emph{is correct not only in the} $\gamma_{0}$ -
\emph{frame with the standard basis} $\left\{  \gamma_{\mu}\right\}  $
\emph{but in all other relatively moving frames and it holds for any
permissible choice of coordinates, i.e., basis} $\left\{  e_{\mu}\right\}  $.
We see that the relativistically correct fields $E$ and $B$ and the new field
equations (\ref{deb}) and (\ref{maeb}) do not have the same physical
interpretation as the usual 3D fields $\mathbf{E}$ and $\mathbf{B}$ and the
usual 3D ME (\ref{MEC}) except in the $\gamma_{0}$ - frame with the $\left\{
\gamma_{\mu}\right\}  $ basis in which $E^{0}=B^{0}=0$. This consideration
completely defines the relation between our approach with 4D $E$ and $B$ and
all previous approaches. \bigskip\medskip

\textbf{III.} \textbf{THE PROOF IN THE GEOMETRIC\ ALGEBRA FORMALISM}

\textbf{USING\ BIVECTORS\ }$\mathbf{E}_{H}$ \textbf{AND} $\mathbf{B}_{H}%
$\textbf{\ }\bigskip

\textbf{A. The field equations in the} $\gamma_{0}$ - \textbf{frame.}
\textbf{The Maxwell equations\bigskip}

The same proof and the whole consideration as in section II. can be repeated
using in the $\gamma_{0}$ - frame with the $\left\{  \gamma_{\mu}\right\}  $
basis the decomposition of $F$ into the bivectors $\mathbf{E}_{H}$ and
$\mathbf{B}_{H}$ (\ref{FB}) instead of the decomposition of $F$ into 1-vectors
$E_{f}$ and $B_{f}$ (\ref{ebg}). It will be seen that the type of the algebric
object chosen to represent the electric and magnetic fields is irrelevant for
the whole consideration and for the obtained results. We shall briefly repeat
the main results from section II. but starting with $\mathbf{E}_{H}$ and
$\mathbf{B}_{H}$ instead of $E_{f}$ and $B_{f}$. When the decomposition
(\ref{FB}) is substituted into the field equations (\ref{MEF}) we find
\begin{align}
\partial\lbrack(F\cdot\gamma_{0})\wedge\gamma_{0}+(F\wedge\gamma_{0}%
)\cdot\gamma_{0}]  &  =j/\varepsilon_{0}c,\nonumber\\
\partial(\mathbf{E}_{H}+c\gamma_{5}\mathbf{B}_{H})  &  =j/\varepsilon_{0}c.
\label{H1}%
\end{align}
All quantities in (\ref{H1}) can be written as CBGQs in the standard basis
$\left\{  \gamma_{\mu}\right\}  $, see [6],
\begin{equation}
\mathbf{E}_{H}=F^{i0}\gamma_{i}\wedge\gamma_{0},\quad\mathbf{B}_{H}%
=(1/2c)\varepsilon^{kli0}F_{kl}\gamma_{i}\wedge\gamma_{0}. \label{aj}%
\end{equation}
It is seen from (\ref{aj}) that both bivectors $\mathbf{E}_{H}$ and
$\mathbf{B}_{H}$ are parallel to $\gamma_{0}$, that is, it holds that
$\mathbf{E}_{H}\wedge\gamma_{0}=\mathbf{B}_{H}\wedge\gamma_{0}=0$. Further it
follows from (\ref{aj}) that the components of $\mathbf{E}_{H},$
$\mathbf{B}_{H}$ in the $\left\{  \gamma_{\mu}\right\}  $ basis give rise to
the tensor (components)$\;(\mathbf{E}_{H})^{\mu\nu}=\gamma^{\nu}\cdot
(\gamma^{\mu}\cdot\mathbf{E}_{H})=(\gamma^{\nu}\wedge\gamma^{\mu}%
)\cdot\mathbf{E}_{H},$ (and the same for $(\mathbf{B}_{H})^{\mu\nu}$) which,
written out as a matrix, have entries
\begin{align}
(\mathbf{E}_{H})^{i0}  &  =F^{i0}=-(\mathbf{E}_{H})^{0i}=E^{i},\quad
(\mathbf{E}_{H})^{ij}=0,\nonumber\\
(\mathbf{B}_{H})^{i0}  &  =(1/2c)\varepsilon^{kli0}F_{kl}=-(\mathbf{B}%
_{H})^{0i}=B^{i},\quad(\mathbf{B}_{H})^{ij}=0. \label{ad}%
\end{align}
Then (\ref{aj}) becomes
\begin{align}
\mathbf{E}_{H}  &  =(E_{H})^{i0}\gamma_{i}\wedge\gamma_{0}=E^{i}\gamma
_{i}\wedge\gamma_{0},\nonumber\\
\mathbf{B}_{H}  &  =(B_{H})^{i0}\gamma_{i}\wedge\gamma_{0}=B^{i}\gamma
_{i}\wedge\gamma_{0}. \label{aj1}%
\end{align}
\emph{Multiplying} (\ref{H1}) \emph{by} $\gamma_{0}$ and using (\ref{aj}) and
(\ref{ad}) we write the resulting equations as a coordinate-based geometric
equation
\begin{align}
(\partial_{k}E^{k}-j^{0}/c\varepsilon_{0})+(\partial_{0}E^{i}-c\varepsilon
^{ijk0}\partial_{j}B_{k}+j^{i}/c\varepsilon_{0})(\gamma_{i}\wedge\gamma_{0})+
& \nonumber\\
(c\partial_{k}B^{k})\gamma_{5}+(c\partial_{0}B^{i}+\varepsilon^{ijk0}%
\partial_{j}E_{k})\gamma_{5}(\gamma_{i}\wedge\gamma_{0})  &  =0. \label{H2}%
\end{align}
The equation (\ref{H2}) is exactly the same as the equations obtained in the
standard geometric algebra formalism, e.g., (8.5) and (8.6a-8.6d) in [7]
Space-Time Algebra,\textit{\ }but now written as a coordinate-based geometric
equation. (\ref{H2}) encodes all four ME in the component form in the same way
as it happens with the equation (\ref{MEC}). It is worth noting that this
step, the multiplication of (\ref{H1}) by $\gamma_{0}$, in order to get the
usual ME, is unnecessary in the formulation from section II. with 1-vectors
$E_{f}$ and $B_{f}.$ This shows that the approach with 1-vectors $E_{f}$ and
$B_{f}$ is simpler than the approach with bivectors $\mathbf{E}_{H}$ and
$\mathbf{B}_{H}$ and also it is much closer to the classical formulation of
the electromagnetism with the 3D vectors $\mathbf{E}$ and $\mathbf{B}%
.$\medskip\bigskip

\textbf{B. Lorentz transformations of the Maxwell equations} \bigskip

Let us now apply the active LT (using (\ref{err})) to (\ref{H2}). First we
rewrite (\ref{H2}) in the form
\begin{equation}
a^{0}+a^{i}(\gamma_{i}\wedge\gamma_{0})+b^{0}\gamma_{5}+b^{i}\gamma_{5}%
(\gamma_{i}\wedge\gamma_{0})=0. \label{H3}%
\end{equation}
The coefficients $a^{0}$, $a^{i}$ and $b^{0}$, $b^{i}$ are clear from
(\ref{H2}); they are the usual ME in the component form. As it is said the
usual ME (\ref{H2}), i.e., (\ref{H3}), are obtained multiplying the equations
(\ref{H1}) by $\gamma_{0}$ The LT of the resulting equations (after
multiplication by $\gamma_{0}$) are $R\{\gamma_{0}[\partial((F\cdot\gamma
_{0})\wedge\gamma_{0}+(F\wedge\gamma_{0})\cdot\gamma_{0})-j/\varepsilon
_{0}c]\}\widetilde{R}=0,$ that is,
\begin{equation}
R\{\gamma_{0}[\partial(\mathbf{E}_{H}+c\gamma_{5}\mathbf{B}_{H})-j/\varepsilon
_{0}c]\}\widetilde{R}=0. \label{reg}%
\end{equation}
Then after applying the LT upon (\ref{H2}), i.e., (\ref{H3}), we find
\begin{equation}
a^{0}+R[a^{i}(\gamma_{i}\wedge\gamma_{0})]\widetilde{R}+b^{0}\gamma
_{5}+R[b^{i}\gamma_{5}(\gamma_{i}\wedge\gamma_{0})]\widetilde{R}=0, \label{H4}%
\end{equation}
where $R[a^{i}(\gamma_{i}\wedge\gamma_{0})]\widetilde{R}=a^{1}(\gamma
_{1}\wedge\gamma_{0})+\gamma\lbrack a^{2}(\gamma_{2}\wedge\gamma_{0}%
)+a^{3}(\gamma_{3}\wedge\gamma_{0})]-\beta\gamma\lbrack a^{2}(\gamma_{2}%
\wedge\gamma_{1})+a^{3}(\gamma_{3}\wedge\gamma_{1})]$ and $R[b^{i}\gamma
_{5}(\gamma_{i}\wedge\gamma_{0})]\widetilde{R}=b^{1}(\gamma_{3}\wedge
\gamma_{2})+\gamma\lbrack-b^{2}(\gamma_{3}\wedge\gamma_{1})+b^{3}(\gamma
_{2}\wedge\gamma_{1})]+\beta\gamma\lbrack b^{2}(\gamma_{3}\wedge\gamma
_{0})-b^{3}(\gamma_{2}\wedge\gamma_{0})]$. This result (\ref{H4}) is the usual
result for the active LT of a multivector from (\ref{H3}).

The above equation (\ref{reg}) can be expressed in terms of Lorentz
transformed derivatives and Lorentz transformed $\mathbf{E}_{H}^{\prime}$ and
$\mathbf{B}_{H}^{\prime}$ as
\begin{equation}
\gamma_{0}^{\prime}[\partial^{\prime}(\mathbf{E}_{H}^{\prime}+c\gamma
_{5}\mathbf{B}_{H}^{\prime})-j^{\prime}/\varepsilon_{0}c]=0, \label{ehbc}%
\end{equation}
where $\gamma_{0}^{\prime}=R\gamma_{0}\widetilde{R}$, $\partial^{\prime
}=R\partial\widetilde{R}$, and (see also [6]) the Lorentz transformed
bivectors are $\mathbf{E}_{H}^{\prime}$ and $\mathbf{B}_{H}^{\prime}$. This
$\mathbf{E}_{H}^{\prime}$ is%

\begin{align}
\mathbf{E}_{H}^{\prime}  &  =R[(F\cdot\gamma_{0})\gamma_{0}]\widetilde
{R}=R\mathbf{E}_{H}\widetilde{R}=E^{1}\gamma_{1}\wedge\gamma_{0}+\gamma
(E^{2}\gamma_{2}\wedge\gamma_{0}+\nonumber\\
&  E^{3}\gamma_{3}\wedge\gamma_{0})-\beta\gamma(E^{2}\gamma_{2}\wedge
\gamma_{1}+E^{3}\gamma_{3}\wedge\gamma_{1}). \label{eh}%
\end{align}
The components $(\mathbf{E}_{H}^{\prime})^{\mu\nu}$ that are different from
zero are $(\mathbf{E}_{H}^{\prime})^{10}=E^{1}$, $(\mathbf{E}_{H}^{\prime
})^{20}=\gamma E^{2},$ $(\mathbf{E}_{H}^{\prime})^{30}=\gamma E^{\text{3}},$
$(\mathbf{E}_{H}^{\prime})^{12}=\beta\gamma E^{2}$, $(\mathbf{E}_{H}^{\prime
})^{13}=\beta\gamma E^{3}$. $(\mathbf{E}_{H}^{\prime})^{\mu\nu}$ is
antisymmetric, i.e., $(\mathbf{E}_{H}^{\prime})^{\nu\mu}=-(\mathbf{E}%
_{H}^{\prime})^{\mu\nu}$ and we denoted, as in (\ref{ad}), $E^{i}=F^{i0}$.
Similarly we find for $\mathbf{B}_{H}^{\prime}$
\begin{align}
\mathbf{B}_{H}^{\prime}  &  =R[(-1/c)\gamma_{5}((F\wedge\gamma_{0})\cdot
\gamma_{0})]\widetilde{R}=R\mathbf{B}_{H}\widetilde{R}=B^{1}\gamma_{1}%
\wedge\gamma_{0}+\nonumber\\
&  \gamma(B^{2}\gamma_{2}\wedge\gamma_{0}+B^{3}\gamma_{3}\wedge\gamma
_{0})-\beta\gamma(B^{2}\gamma_{2}\wedge\gamma_{1}+B^{3}\gamma_{3}\wedge
\gamma_{1}). \label{Be}%
\end{align}
The components $(\mathbf{B}_{H}^{\prime})^{\mu\nu}$ that are different from
zero are $(\mathbf{B}_{H}^{\prime})^{10}=B^{1}$, $(\mathbf{B}_{H}^{\prime
})^{20}=\gamma B^{2},$ $(\mathbf{B}_{H}^{\prime})^{30}=\gamma B^{\text{3}},$
$(\mathbf{B}_{H}^{\prime})^{12}=\beta\gamma B^{2}$, $(\mathbf{B}_{H}^{\prime
})^{13}=\beta\gamma B^{3}$. $(\mathbf{B}_{H}^{\prime})^{\mu\nu}$ is
antisymmetric, i.e., $(\mathbf{B}_{H}^{\prime})^{\nu\mu}=-(\mathbf{B}%
_{H}^{\prime})^{\mu\nu}$ and we denoted, as in (\ref{ad}), $B^{i}%
=(1/2c)\varepsilon^{kli0}F_{kl}$. Both (\ref{eh}) and (\ref{Be}) are the
familiar forms for the active LT of bivectors, here $\mathbf{E}_{H}$ and
$\mathbf{B}_{H}$. It is worth noting that $\mathbf{E}_{H}^{\prime}$ and
$\mathbf{B}_{H}^{\prime}$, in contrast to $\mathbf{E}_{H}$ and $\mathbf{B}%
_{H}$, are not parallel to $\gamma_{0}$, i.e., it \emph{does not hold}
\emph{that} $\mathbf{E}_{H}^{\prime}\wedge\gamma_{0}=\mathbf{B}_{H}^{\prime
}\wedge\gamma_{0}=0$ and thus \emph{there are} $(\mathbf{E}_{H}^{\prime}%
)^{ij}\neq0$ \emph{and} $(\mathbf{B}_{H}^{\prime})^{ij}\neq0.$ Further, as it
happens for $E_{f}$ and $B_{f}$, see (\ref{nle}) and (\ref{nlb}), \emph{the
components} $(\mathbf{E}_{H})^{\mu\nu}$ ($(\mathbf{B}_{H})^{\mu\nu}$)
\emph{transform upon the active LT again to the components} $(\mathbf{E}%
_{H}^{\prime})^{\mu\nu}$ ($(\mathbf{B}_{H}^{\prime})^{\mu\nu}$); \emph{there
is no mixing of components}. \emph{Thus} \emph{by the active LT}
$\mathbf{E}_{H}$ \emph{transforms to} $\mathbf{E}_{H}^{\prime}$ \emph{and}
$\mathbf{B}_{H}$ \emph{to }$\mathbf{B}_{H}^{\prime}.$ Actually, as we said,
this is the way in which every bivector transforms upon the active LT. The
last form of the Lorentz transformed field equation, (\ref{ehbc}), can be
written as a coordinate-based geometric equation in the standard basis
$\left\{  \gamma_{\mu}\right\}  $, but for simplicity we only quote the scalar
term $a^{\prime0}$
\begin{align}
a^{\prime0}  &  =-\beta\gamma\partial_{0}^{\prime}(\mathbf{E}_{H}^{\prime
})^{10}+\gamma\lbrack\partial_{k}^{\prime}(\mathbf{E}_{H}^{\prime}%
)^{k0}]+\beta\gamma\lbrack\partial_{2}^{\prime}(\mathbf{E}_{H}^{\prime}%
)^{21}\nonumber\\
&  +\partial_{3}^{\prime}(\mathbf{E}_{H}^{\prime})^{31}]-(\gamma j^{\prime
0}-\beta\gamma j^{\prime1})/\varepsilon_{0}c \label{act}%
\end{align}
Comparing $a^{\prime0}$ (\ref{act}) with $a^{0}$ from the usual ME (\ref{H2}),
i.e., (\ref{H3}), $a^{0}=\partial_{k}(\mathbf{E}_{H})^{k0}-j^{0}%
/c\varepsilon_{0}$, we again see, as in section II.B with $E_{f}$ and $B_{f}$,
that $a^{\prime0}$ \emph{substantially differs in form from the term} $a^{0}$
in (\ref{H2}). Again the same situation happens with other transformed terms,
which shows, as in section II.B, that \emph{the Lorentz transformed ME},
(\ref{ehbc}) \emph{with} (\ref{act}), \emph{are not of the same form as the
original ones} (\ref{H2}), i.e., (\ref{H3}). This is a fundamental result
which once again reveals that, contrary to the previous derivations, e.g.,
[2,16], [3,4], [7-9], and contrary to the generally accepted belief, \emph{the
usual ME are not Lorentz covariant equations.\bigskip\medskip}

\textbf{C. Standard transformations of the Maxwell equations} \emph{\bigskip}

As can be easily shown, see also [6], \emph{the ST for} $\mathbf{E}%
_{H,st}^{\prime}$ \emph{and} $\mathbf{B}_{H,st}^{\prime}$ \emph{are derived
wrongly assuming that the quantities obtained by the active LT of}
$\mathbf{E}_{H}$ \emph{and} $\mathbf{B}_{H}$ \emph{are again parallel to}
$\gamma_{0}$\emph{, i.e., that again holds} $\mathbf{E}_{H}^{\prime}%
\wedge\gamma_{0}=\mathbf{B}_{H}^{\prime}\wedge\gamma_{0}=0$ and consequently
that $(\mathbf{E}_{H,st}^{\prime})^{ij}=(\mathbf{B}_{H,st}^{\prime})^{ij}=0.$
Thence, in contrast to the correct LT of $\mathbf{E}_{H}$ \emph{and}
$\mathbf{B}_{H},$ (\ref{eh}) and (\ref{Be}) respectively, it is taken in
standard derivations ([7], Space-Time Algebra (eq. (18.22)), New Foundations
for Classical Mechanics (Ch. 9 eqs. (3.51a,b)), [8] Geometric algebra for
physicists (Ch. 7.1.2 eq. (7.33))) that
\begin{align}
\mathbf{E}_{H,st}^{\prime}  &  =(F^{\prime}\cdot\gamma_{0})\gamma
_{0}=(E_{H,st}^{\prime})^{i0}\gamma_{i}\wedge\gamma_{0}=E_{st}^{\prime
i}\gamma_{i}\wedge\gamma_{0}=\nonumber\\
&  E^{1}\gamma_{1}\wedge\gamma_{0}+(\gamma E^{2}-\beta\gamma cB^{3})\gamma
_{2}\wedge\gamma_{0}+(\gamma E^{3}+\beta\gamma cB^{2})\gamma_{3}\wedge
\gamma_{0}, \label{es}%
\end{align}
where $F^{\prime}=RF\widetilde{R}$. Similarly we find for $\mathbf{B}%
_{H,st}^{\prime}$
\begin{align}
\mathbf{B}_{H,st}^{\prime}  &  =(-1/c)\gamma_{5}[(F^{\prime}\wedge\gamma
_{0})\cdot\gamma_{0})]=(B_{H,st}^{\prime})^{i0}\gamma_{i}\wedge\gamma
_{0}=B_{st}^{\prime i}\gamma_{i}\wedge\gamma_{0}=\nonumber\\
&  B^{1}\gamma_{1}\wedge\gamma_{0}+(\gamma B^{2}+\beta\gamma E^{3}%
/c)\gamma_{2}\wedge\gamma_{0}+(\gamma B^{3}-\beta\gamma E^{2}/c)\gamma
_{3}\wedge\gamma_{0}. \label{bes}%
\end{align}
The relations (\ref{es}) and (\ref{bes}) immediately give the familiar
expressions for the ST of the 3D vectors $\mathbf{E}$ and $\mathbf{B.}$ Now,
in contrast to the correct LT of $\mathbf{E}_{H}$ \emph{and} $\mathbf{B}_{H},$
(\ref{eh}) and (\ref{Be}) respectively, \emph{the components} \emph{of the
transformed }$\mathbf{E}_{H,st}^{\prime}$ \emph{are expressed by the mixture
of components} $E^{i}$ \emph{and} $B^{i},$ \emph{and the same holds for}
$\mathbf{B}_{H,st}^{\prime}$.

Here we again explicitly show that in the standard derivations [7-9] the ME
(\ref{H2}) remain unchanged in form not upon the LT but upon some
transformations which, strictly speaking, have nothing to do with the LT of
the equation (\ref{H2}). Namely the ST of the second equation in (\ref{H1})
(after multiplication by $\gamma_{0}$) are given as
\begin{equation}
\gamma_{0}[\partial^{\prime}(\mathbf{E}_{H,st}^{\prime}+c\gamma_{5}%
\mathbf{B}_{H,st}^{\prime})-j^{\prime}/\varepsilon_{0}c]=0, \label{gc}%
\end{equation}
where $\mathbf{E}_{H,st}^{\prime}$ and $\mathbf{B}_{H,st}^{\prime}$ are
determined by (\ref{es}) and (\ref{bes}). Notice again that, in contrast to
the correct LT (\ref{reg}) or (\ref{ehbc}), $\gamma_{0}$ \emph{is not
transformed in} (\ref{gc}), as it is not transformed in the ST of the electric
and magnetic fields (\ref{es}) and (\ref{bes}). When (\ref{gc}) is written as
a coordinate-based geometric equation in the standard basis $\left\{
\gamma_{\mu}\right\}  $ it becomes
\begin{align}
(\partial_{k}^{\prime}E_{st}^{\prime k}-j^{\prime0}/c\varepsilon
_{0})+(\partial_{0}^{\prime}E_{st}^{\prime i}-c\varepsilon^{ijk0}\partial
_{j}^{\prime}B_{st,k}^{\prime}+j^{\prime i}/c\varepsilon_{0})(\gamma_{i}%
\wedge\gamma_{0})+  & \nonumber\\
(c\partial_{k}^{\prime}B_{st}^{\prime k})\gamma_{5}+(c\partial_{0}^{\prime
}B_{st}^{\prime i}+\varepsilon^{ijk0}\partial_{j}^{\prime}E_{st,k}^{\prime
})\gamma_{5}(\gamma_{i}\wedge\gamma_{0})  &  =0. \label{meq}%
\end{align}

The equation (\ref{meq}) is of the same form as the original ME (\ref{H2}) but
the electric and magnetic fields are not transformed by the LT than by the ST.
As seen from (\ref{gc}) (together with (\ref{es}) and (\ref{bes})) \emph{the
equation (\ref{meq}) is not the LT of the original ME} (\ref{H2}); \emph{the
LT of the ME} (\ref{H2}) \emph{is the equation} (\ref{ehbc}) \emph{with}
(\ref{act}) \emph{(i.e.,} (\ref{reg}) or (\ref{H4})), where the Lorentz
transformed electric and magnetic fields are given by the relations (\ref{eh})
and (\ref{Be}).\medskip\bigskip

\textbf{D. Lorentz\ invariant\ field\ equations\ with bivectors }$E_{HL}$
\textbf{and }$B_{HL}$\textbf{\ }\bigskip

As explained in the preceding sections the \emph{observer independent} $F$
field is decomposed in (\ref{FB}) (see [7,8]) in terms of \emph{observer
dependent quantities,} i.e., as the sum of a relative vector $\mathbf{E}_{H}$
and a relative bivector $\gamma_{5}\mathbf{B}_{H},$ by making the space-time
split in the $\gamma_{0}$ - frame. But, similarly as in section II.D., we
present here \emph{an observer independent} decomposition of $F$ into
bivectors $E_{HL}$ and $B_{HL}$ that are defined without reference frames,
i.e., \emph{which are independent of the chosen reference frame and of the
chosen system of coordinates in it.} We define
\begin{align}
F  &  =E_{HL}+c\gamma_{5}B_{HL}\mathbf{,\quad} E_{HL}=(1/c^{2})(F\cdot
v)\wedge v\nonumber\\
B_{HL}  &  =-(1/c^{3})\gamma_{5}[(F\wedge v)\cdot v],\quad\gamma_{5}%
B_{HL}=(1/c^{3})(F\wedge v)\cdot v \label{he}%
\end{align}
(The subscript 'HL' is for - Hestenes, Lasenby, see [21].) Of course, as in
II.D., \emph{the velocity} $v$ \emph{and all other quantities entering into}
(\ref{he}) \emph{are defined without reference frames.} Consequently
(\ref{he}) \emph{holds for any observer.} When (\ref{he}) is used the field
equation for $F$ (\ref{MEF}), after multiplication by $v/c$ (instead of by
$\gamma_{0}$), becomes
\begin{equation}
(v/c)\{\partial(E_{HL}+c\gamma_{5}B_{HL})-j/\varepsilon_{0}c\}=0. \label{Nf}%
\end{equation}
In contrast to the field equation (\ref{H1}) that holds only for the
$\gamma_{0}$-observer, the field equation (\ref{Nf}) \emph{holds for any
observer;} \emph{the quantities entering into} (\ref{Nf}) \emph{are all
defined without reference frames.} \emph{The equation} (\ref{Nf}) \emph{is
physicaly completely equivalent to the field equation for} $F$ (\ref{MEF}%
)\emph{, i.e.,} \emph{to the field equation with 1- vectors} $E$ \emph{and}
$B$ (\ref{deb}). (The equation (\ref{H1}) corresponds to the equation
(\ref{eqfi}), while (\ref{Nf}) corresponds to (\ref{deb}).) The field equation
(\ref{Nf}) can be written as a coordinate-based geometric equation, and it
looks much more complicated than the equation (\ref{maeb}) with 1- vectors $E$
and $B$. We write it (for better comparison) as two equations; the first one
will yield the scalar and bivector parts of (\ref{H2}) when $v/c=\gamma_{0}$.
It is
\begin{align}
&  (1/c)v_{\beta}\partial_{\alpha}(E_{HL})^{\alpha\beta}+[(1/2c)v^{\alpha
}\partial_{\alpha}(E_{HL})^{\beta\sigma}-(1/2)\varepsilon^{\mu\nu\alpha\sigma
}v^{\beta}\partial_{\alpha}(B_{HL})_{\mu\nu}]\gamma_{\beta}\wedge
\gamma_{\sigma}\nonumber\\
&  =(1/\varepsilon_{0}c^{2})(v_{\alpha}j^{\alpha}+v^{\beta}j^{\sigma}%
\gamma_{\beta}\wedge\gamma_{\sigma}). \label{nh}%
\end{align}
The second equation will yield the pseudoscalar and pseudobivector parts of
(\ref{H2}) when $v/c=\gamma_{0}$ and it is
\begin{equation}
v_{\beta}\partial_{\alpha}(B_{HL})^{\alpha\beta}\gamma_{5}+(1/2)v^{\alpha
}\partial_{\alpha}(B_{HL})^{\mu\nu}\gamma_{5}(\gamma_{\mu}\wedge\gamma_{\nu
})+(v_{\beta}\partial^{\alpha}-v^{\alpha}\partial_{\beta})(E_{HL})_{\alpha\nu
}\gamma^{\beta}\wedge\gamma^{\nu}=0. \label{D}%
\end{equation}
The equation (\ref{nh}) is with sources and it emerges from $\partial\cdot
F=j/\varepsilon_{0}c$, while (\ref{D}) is the source-free equation and it
emerges from $\partial\wedge F=0$. Comparing (\ref{nh}) and (\ref{D}) in the
$E_{HL}$, $B_{HL}$ - formulation with the corresponding parts in (\ref{maeb})
with 1- vectors $E$ and $B$ we see that the formulation with $E$ and $B$ is
much simpler and more elegant than the formulation with bivectors $E_{HL}$ and
$B_{HL}$; the physical content is completely equivalent.

The equations (\ref{nh}) and (\ref{D}) are written in a manifestly covariant
form. This means that \emph{when the active LT are applied upon such}
(\ref{nh}) \emph{and} (\ref{D}) \emph{the equations remain of the same form
but with primed quantities replacing the unprimed ones (of course the basis is
unchanged). }

The whole discussion from section II.D. (with 1- vectors $E$ and $B$) about
the correspondence principle applies in the same measure to the formulation
with bivectors $E_{HL}$ and $B_{HL}$. The only difference is the simplicity of
the formulation with 1- vectors $E$ and $B$.

The same conclusions hold for the formulation with 1-vector $\mathbf{E}_{J}$
and a bivector $\mathbf{B}_{J}$ from [9], but for the sake of brevity that
formulation will not be considered here. \bigskip\medskip

\textbf{IV. THE PROOF IN THE TENSOR FORMALISM USING\ }

\textbf{4-VECTORS\ }$E^{a}$\textbf{\ AND\ }$B^{a}$\bigskip

The same proof that the classical electromagnetism and SR are not in agreement
can be given in the tensor formalism as well. The important parts of this
issue are already treated in two papers, [11] and [5].

Let us start with some general definitions. The electromagnetic field tensor
$F^{ab}$ is defined without reference frames, i.e., it is an abstract tensor,
a geometric quantity; Latin indices a,b,c, are to be read according to the
abstract index notation, as in [22] and [11,12], [18]. When some reference
frame (a physical object) is introduced and the system of coordinates (a
mathematical object) is adopted in it, then $F^{ab}$ can be written as a CBGQ
containing components and a basis. As already said in the invariant
formulation of SR that uses 4D quantities defined without reference frames
[11,12], [18] and [5] in the tensor formalism, and [10,15] and [6] in the
Clifford algebra formalism, any permissible system of coordinates, not
necessary the Einstein system of coordinates, i.e., the standard basis
$\left\{  \gamma_{\mu}\right\}  $, can be used on an equal footing. However,
for simplicity, in this part we shall deal only with the standard basis
$\left\{  \gamma_{\mu}\right\}  $. When $F^{ab}$ is written as a CBGQ it
becomes $F^{ab}=F^{\mu\nu}\gamma_{\mu}\otimes\gamma_{\nu},$ where Greek
indices $\mu,\nu$ in $F^{\mu\nu}$ run from 0 to 3 and they denote the
components of the geometric object $F^{ab}$ in some system of coordinates,
here the standard basis $\left\{  \gamma_{\mu}\right\}  $. In the tensor
formalism $\gamma_{\mu}$ denote the basis 4-vectors (not components) forming
the standard basis $\left\{  \gamma_{\mu}\right\}  $ and $^{\prime}%
\otimes^{\prime}$ denotes the tensor product of the basis 4-vectors. In the
tensor formalism I shall often denote the unit 4-vector in the time direction
$\gamma_{0}$ as $t^{b}$ as well. Then in some reference frame with the
standard basis $\left\{  \gamma_{\mu}\right\}  $ $t^{b}$ can be also written
as a CBGQ, $t^{b}=t^{\mu}\gamma_{\mu}$, where $t^{\mu}$ is a set of components
of the unit 4-vector in the time direction ($t^{\mu}=$($1,0,0,0$)). Almost
always in the standard covariant approaches to SR one considers only the
components of the geometric quantities taken in the $\left\{  \gamma_{\mu
}\right\}  $ basis and thus not the whole tensor. However the components are
coordinate quantities and they do not contain the whole information about the
physical quantity.\bigskip\medskip

\textbf{A. The field equations in the} $\gamma_{0}$ - \textbf{frame.}
\textbf{The Maxwell equations}\bigskip

In the abstract index notation the field equations are given as
\begin{equation}
(-g)^{-1/2}\partial_{a}((-g)^{1/2}F^{ab})=j^{b}/\varepsilon_{0}c,\quad
\varepsilon^{abcd}\partial_{b}F_{cd}=0 \label{maxten}%
\end{equation}
where $g$ is the determinant of the metric tensor $g_{ab}$ and $\partial_{a}$
is an ordinary derivative operator. Now there are two field equations while in
the geometric algebra formalism they are united in only one field equation.
When written in the $\left\{  \gamma_{\mu}\right\}  $ basis as
coordinate-based geometric equations the relations (\ref{maxten}) become
\begin{equation}
\partial_{\alpha}F^{a\beta}\gamma_{\beta}=(1/\varepsilon_{0}c)j^{\beta}%
\gamma_{\beta},\quad\partial_{\alpha}\ ^{\ast}F^{\alpha\beta}\gamma_{\beta}=0.
\label{maco1}%
\end{equation}
Notice that from (\ref{maco1}) one simply finds the usual covariant form (the
component form in the $\left\{  \gamma_{\mu}\right\}  $ basis) of the field
equations with $F^{\alpha\beta}$ and its dual $^{\ast}F^{\alpha\beta}$
\begin{equation}
\partial_{\alpha}F^{a\beta}=j^{\beta}/\varepsilon_{0}c,\quad\partial_{\alpha
}\ ^{\ast}F^{\alpha\beta}=0, \label{maxco}%
\end{equation}
where $^{\ast}F^{\alpha\beta}=(1/2)\varepsilon^{\alpha\beta\gamma\delta
}F_{\gamma\delta}$. In analogy with the geometric algebra formalism, $F^{ab}$
can be decomposed in terms of the observer dependent 4-vectors $E_{f}^{a}$ and
$B_{f}^{a}$ \emph{by singling out a particular time-like direction }$t^{b}$.
(This corresponds to the decomposition of $F$ into 1-vectors $E_{f}$ and
$B_{f}$ (\ref{ebg}).) Thus
\begin{align}
F^{ab}  &  =\delta_{\quad cd}^{ab}E_{f}^{c}t^{d}+c\varepsilon^{abcd}%
t_{c}B_{f,d},\nonumber\\
E_{f}^{a}  &  =F^{ab}t_{b},\quad B_{f}^{a}=(1/2c)\varepsilon^{abcd}t_{b}%
F_{cd}. \label{veef}%
\end{align}
All quantities from (\ref{veef}) can be written as CBGQs in the standard basis
$\left\{  \gamma_{\mu}\right\}  .$ Then in the tensor formalism we find the
same equations as the equations (\ref{gnl}) in the geometric algebra formalism
with 1-vectors $E_{f}$ and $B_{f}.$ They are
\begin{align}
E_{f}^{a}  &  =E_{f}^{\mu}\gamma_{\mu}=0\gamma_{0}+F^{k0}\gamma_{k}%
,\nonumber\\
B_{f}^{a}  &  =B_{f}^{\mu}\gamma_{\mu}=0\gamma_{0}+(-1/2c)\varepsilon
^{0kli}F_{kl}\gamma_{i}, \label{gt}%
\end{align}
whence we get the relation (\ref{sko}) $E_{f}^{i}=F^{i0},\quad B_{f}%
^{i}=(-1/2c)\varepsilon^{0kli}F_{kl},$ which is, as already said, nothing else
than the standard identification of the components $F^{\mu\nu}$ with the
components of the 3D vectors $\mathbf{E}$ and $\mathbf{B,}$ see, e.g., [16],
[3,4]. (As mentioned previously Einstein's fundamental work [16] is the
earliest reference on generally covariant electrodynamics and on the
identification of some components of $F^{ab}$ (actually $F^{\alpha\beta}$)
with the components of $\mathbf{E}$ and $\mathbf{B.}$ He introduces an
electromagnetic potential 4-vector (in component form) and from this
constructs $F^{a\beta},$ the component form of the $F^{ab}$ tensor. Then he
writes the equations (\ref{maxco}) and shows that these equations correspond
to the usual Maxwell equations with $\mathbf{E}$ and $\mathbf{B}$ if he makes
the identification given in the equations (\ref{sko}). It has to be mentioned
that Einstein actually worked with \emph{the equations for basis components in
the }$\left\{  \gamma_{\mu}\right\}  $ \emph{basis} and thus not with the
abstract tensors, defined without reference frames, or with coordinate-based
geometric equations (see, e.g., [23] for the comparison of Einstein's view of
spacetime and the modern view).) In fact, the whole discussion in connection
with the relations (\ref{gnl}) and (\ref{sko}) applies in the same measure to
(\ref{gt}). Thus in the rest frame of 'fiducial' observers (we again call that
frame - the $\gamma_{0}$ - frame) $E_{f}^{a}$ \emph{and} $B_{f}^{a}$ \emph{do
not have the temporal components} $E_{f}^{0}=B_{f}^{0}=0$; in the $\gamma_{0}$
- frame $t^{\mu}$ can be interpreted as the 4-velocity (the components in the
$\left\{  \gamma_{\mu}\right\}  $ basis) of the observers that are at rest
there. In the standard treatments the 3-vectors $\mathbf{E}$ and $\mathbf{B}$,
as \emph{geometric quantities in the 3D space}, are constructed from the
spatial components $E^{i}$ and $B^{i}$ from (\ref{gt}), i.e., (\ref{sko}), and
\emph{the unit 3-vectors} $\mathbf{i},$ $\mathbf{j},$ $\mathbf{k,}$ e.g.,
$\mathbf{E=}F^{10}\mathbf{i}+F^{20}\mathbf{j}+F^{30}\mathbf{k.}$ These results
are quoted in numerous textbooks and papers treating relativistic
electrodynamics in the tensor formalism, see, e.g., [16], [3,4]. Actually in
the usual covariant approaches, e.g., [16], [3,4], one forgets about $E^{0}$
and $B^{0}$ components and simply makes the identification of six independent
components of $F^{\mu\nu}$ with three components $E^{i}$, $E^{i}=F^{i0}$, and
three components $B^{i}$, $B^{i}=(1/2)\varepsilon^{ikl}F_{lk}.$ Since in SR we
work with the 4D spacetime the mapping between the components of $F^{\mu\nu}$
and the components of the 3D vectors $\mathbf{E}$ and $\mathbf{B}$ is
mathematically better founded by the relations (\ref{gt}) than by their simple
identification. Therefore we proceed the consideration using (\ref{gt}). Note
again that the whole procedure is made in an inertial frame of reference with
the standard basis $\left\{  \gamma_{\mu}\right\}  $. In another system of
coordinates that is different than the Einstein system of coordinates, e.g.,
differing in the chosen synchronization (as it is the 'r' synchronization
considered in [11]), the identification of $E^{i}$ with $F^{i0},$ as in
(\ref{gt}), i.e., (\ref{sko}), (and also for $B^{i}$), is impossible and
meaningless. Further the components $E^{i}$ and $B^{i}$ are determined in the
4D spacetime in the standard basis $\left\{  \gamma_{\mu}\right\}  .$ Thence
when forming the geometric quantities the components would need to be
multiplied with the unit 4-vectors $\gamma_{i}$ and not with the unit 3-vectors.

Substituting (\ref{veef}) (but written in the $\left\{  \gamma_{\mu}\right\}
$ basis, where $F^{\alpha\beta}=\delta_{\quad\mu\nu}^{\alpha\beta}E_{f}^{\mu
}t^{\nu}+c\varepsilon^{\alpha\beta\mu\nu}t_{\mu}B_{f,\nu}$) into (\ref{maco1})
we find the coordinate-based geometric equations with $E_{f}^{\mu}$,
$B_{f}^{\mu}$ and $t^{\nu}$ as
\begin{align}
\partial_{\alpha}(\delta_{\quad\mu\nu}^{\alpha\beta}E_{f}^{\mu}t^{\nu
}+c\varepsilon^{\alpha\beta\mu\nu}t_{\mu}B_{f,\nu})\gamma_{\beta} &
=(j^{\beta}/\varepsilon_{0})\gamma_{\beta}\nonumber\\
\partial_{\alpha}(\delta_{\quad\mu\nu}^{\alpha\beta}t^{\mu}cB_{f}^{\nu
}+\varepsilon^{\alpha\beta\mu\nu}t_{\mu}E_{f,\nu})\gamma_{\beta} &
=0.\label{me1}%
\end{align}
Using (\ref{gt}) and $t^{\alpha}=(1,0,0,0)$ in (\ref{me1}) these equations
become the same equations as (\ref{MEC}), that is, the usual Maxwell equations
in the component form. They are
\begin{align}
(\partial_{k}E_{f}^{k}-j^{0}/c\varepsilon_{0})\gamma_{0}+(-\partial_{0}%
E_{f}^{i}+c\varepsilon^{ijk0}\partial_{j}B_{fk}-j^{i}/c\varepsilon_{0}%
)\gamma_{i} &  =0\nonumber\\
(-c\partial_{k}B_{f}^{k})\gamma_{0}+(c\partial_{0}B_{f}^{i}+\varepsilon
^{ijk0}\partial_{j}E_{fk})\gamma_{i} &  =0.\label{me}%
\end{align}
The same discussion holds for (\ref{me}) ((\ref{me1})) as for (\ref{MEC})
((\ref{cl})).\medskip\bigskip

\textbf{B. Lorentz transformations of the Maxwell equations} \bigskip

Let us now apply the passive LT to the equations (\ref{me1}), or (\ref{me});
in the tensor formalism we shall deal with the passive LT. Upon the passive LT
the sets of components $E_{f}^{\mu}$ and $B_{f}^{\mu}$ determined in the
$\gamma_{0}$ - frame (the $S$ frame) from (\ref{gt}) transform to
$E_{f}^{\prime\mu}$ and $B_{f}^{\prime\mu}$ in the relatively moving IFR
$S^{\prime}$
\begin{align}
E_{f}^{\prime\mu}  &  =F^{\prime\mu\nu}v_{\nu}^{\prime},\quad B_{f}^{\prime
\mu}=(1/2)\varepsilon^{\mu\nu\lambda\sigma}F_{\lambda\sigma}^{\prime}v_{\nu
}^{\prime}=(F^{\ast})^{\prime\mu\nu}v_{\nu}^{\prime},\nonumber\\
E_{f}^{\prime\mu}  &  =\left(  -\beta\gamma E^{1},\gamma E^{1},E^{2}%
,E^{3}\right)  ,\ B_{f}^{\prime\mu}=\left(  -\beta\gamma B^{1},\gamma
B^{1},B^{2},B^{3}\right)  , \label{ebcr}%
\end{align}
where $v_{\nu}^{\prime}=\left(  \gamma,\beta\gamma,0,0\right)  ,$ and $v_{\nu
}^{\prime}$ is not in the time direction in $S^{\prime}$, i.e., it is not
$=t_{\nu}^{\prime}$. The unit 4-vector (the components) $t^{\mu}$ in the time
direction in $S$ transforms upon the LT into the unit 4-vector $v^{\prime\nu}%
$, the 4-velocity of the moving observers, that contains not only the temporal
component but also $\neq0$ spatial components. Thence, the LT transform the
set of components (\ref{gt}) into (\ref{ebcr}). Note that $E_{f}^{\prime\mu}%
$\emph{\ and }$B_{f}^{\prime\mu}$\emph{\ do have the temporal components as
well. }Further \emph{the components }$E_{f}^{\mu}$ ($B_{f}^{\mu}$) \emph{in
}$S$ \emph{transform upon the LT again to the components} $E_{f}^{\prime\mu}$
($B_{f}^{\prime\mu}$) \emph{in} $S^{\prime}$\emph{; there is no mixing of
components.} Actually this is the way in which every well-defined 4-vector
(the components) transforms upon the LT. A geometric quantity, an abstract
tensor $E^{a},$ can be represented by CBGQs in $S$ and $S^{\prime}$ (both with
the Einstein system of coordinates) as $E_{f}^{\mu}\gamma_{\mu}$ and
$E_{f}^{\prime\mu}\gamma_{\mu}^{\prime},$ where $E_{f}^{\mu}$ and
$E_{f}^{\prime\mu}$ are given by the relations (\ref{gt}) and (\ref{ebcr})
respectively. \emph{All the primed quantities (components and the basis) are
obtained from the corresponding unprimed quantities through the LT.} Of course
it must hold that
\begin{equation}
E^{a}=E_{f}^{\mu}\gamma_{\mu}=E_{f}^{\prime\mu}\gamma_{\mu}^{\prime},
\label{ea}%
\end{equation}
since the components $E_{f}^{\mu}$ transform by the LT, while the basis
$\gamma_{\mu}$ transforms by the inverse LT, \emph{thus leaving the whole
CBGQ\ invariant upon the passive LT}. The invariance of some 4D CBGQ\ upon the
passive LT is the crucial requirement that must be satisfied by any
well-defined 4D quantity. It reflects the fact that such mathematical,
invariant, geometric 4D quantity represents \emph{the same physical object}
for relatively moving observers. The use of CBGQs enables us to have clearly
and correctly defined the concept of sameness of a physical system for
different observers. The importance of this concept in SR was first pointed
out in [24,25]. However they also worked with components in the Einstein
system of coordinates (the covariant quantities) and not with geometric
quantities (the invariant quantities). It is worth noting that in all other
standard treatments, e.g., [2-4] (and [7-9] in the geometric algebra
formalism), the importance of such concept is completely overlooked what
caused many difficulties in understanding SR. It can be easily checked by the
direct inspection that (\ref{ea}) holds when $E_{f}^{\mu}$ and $E_{f}%
^{\prime\mu}$ are given by (\ref{gt}) and (\ref{ebcr}). (The same holds for
$B^{a}.$)

The equations (\ref{me1}), or (\ref{me}), can be written as $a^{\alpha}%
\gamma_{\alpha}=0$ and $b^{\alpha}\gamma_{\alpha}=0$, similarly to the
equation (\ref{ab})$.$ The coefficients $a^{\alpha}$ and $b^{\alpha}$ are
clear from the first and second equation respectively in (\ref{me1}), or
(\ref{me}); \emph{they are the usual Maxwell equations in the component form}.
Then upon the passive LT the equations (\ref{me1}), or (\ref{me}), transform
to
\begin{equation}
a^{\prime\alpha}\gamma_{\alpha}^{\prime}=0,\quad b^{\prime\alpha}%
\gamma_{\alpha}^{\prime}=0, \label{abc}%
\end{equation}
and it holds, as for any 4-vector (a geometric quantity), that $a^{\prime
\alpha}\gamma_{\alpha}^{\prime}=a^{\alpha}\gamma_{\alpha},$ and $b^{\prime
\alpha}\gamma_{\alpha}^{\prime}=b^{\alpha}\gamma_{\alpha}$; the coefficients
transform by the LT as $a^{\prime0}=\gamma a^{0}-\beta\gamma a^{1}$,
$a^{\prime1}=\gamma a^{1}-\beta\gamma a^{0}$, $a^{\prime2}=a^{2}$,
$a^{\prime3}=a^{3}$ (and the same for $b^{\prime\alpha}$), while the basis
4-vectors transform by the inverse LT as $\gamma_{0}^{\prime}=\gamma\gamma
_{0}+\beta\gamma\gamma_{1}$, $\gamma_{1}^{\prime}=\gamma\gamma_{1}+\beta
\gamma\gamma_{0}$, $\gamma_{2}^{\prime}=\gamma_{2}$, $\gamma_{3}^{\prime
}=\gamma_{3}.$ Of course $t^{\nu}$ transforms to $v^{\prime\nu}$ and
$E_{f}^{\prime\mu}$, $B_{f}^{\prime\mu}$ are given by (\ref{ebcr}). (The
equation (\ref{abc}) corresponds to the equation (\ref{L1}) in the geometric
algebra formalism with 1-vectors $E_{f}$ and $B_{f}$.) Again we see that,
e.g., the Gauss law for the electric field $a^{0}$ does not transform by the
LT again to the Gauss law but to $a^{\prime0}$, which is a combination of the
Gauss law and a part of the Amp\`{e}re-Maxwell law ($a^{1}$). When the
coefficients $a^{\prime\alpha}$ and $b^{\prime\alpha}$ are written in terms of
the primed quantities (from the $S^{\prime}$ frame) they become (for
simplicity only the coefficient $a^{\prime0}$ is written)
\begin{equation}
a^{\prime0}=\gamma(\partial_{k}^{\prime}E_{f}^{\prime k})+\beta\gamma
\lbrack\partial_{1}^{\prime}E_{f}^{\prime0}+c(\partial_{2}^{\prime}%
B_{f3}^{\prime}-\partial_{3}^{\prime}B_{f2}^{\prime})]-j^{\prime
0}/c\varepsilon_{0}, \label{ac0}%
\end{equation}
and \emph{it is completely different in form than the coefficient }%
$a^{0}=(\partial_{k}E_{f}^{k}-j^{0}/c\varepsilon_{0})$ in (\ref{me}).
((\ref{ac0}) corresponds to (\ref{anu}).) Again, it can be concluded from
(\ref{ac0}) that the LT do not transform the Gauss law into the 'primed' Gauss
law but into a quite different law; $a^{\prime0}$ contains the time component
$E_{f}^{\prime0}$ while the starting, unprimed $E_{f}^{0}$ is $E_{f}^{0}=0$.
Also the new ''Gauss law'' includes the derivatives of the magnetic field. The
same situation happens with the other Lorentz transformed terms, which once
again explicitly shows that \emph{neither in the tensor formalism the Lorentz
transformed ME} (\ref{abc}) \emph{with} (\ref{ac0}) \emph{are of the same form
as the original ones} (\ref{me}). As discussed in section II.B. this
fundamental result reveals, in the tensor formalism as well, that, contrary to
all previous derivations, e.g., [2-4], and contrary to the generally accepted
opinion, \emph{the usual ME are not Lorentz covariant equations. }\bigskip\medskip

\textbf{C. Standard transformations of the Maxwell equations} \bigskip

In this section we present the derivation of the ST of the ME in the tensor
formalism which is in a complete analogy with the derivation in section II.C..
In all usual treatments, e.g., [3] and [4] eqs. (3.5) and (3.24), in
$S^{\prime}$ one again simply makes the identification of six independent
components of $F^{\prime\mu\nu}$ with three components $E^{\prime i}$,
$E^{\prime i}=F^{\prime i0}$, and three components $B^{\prime i}$, $B^{\prime
i}=(1/2)\varepsilon^{ikl}F_{lk}^{\prime}.$ This means that standard treatments
assume that under the passive LT the set of components $t^{\nu}=\left(
1,0,0,0\right)  $ from $S$ transforms to $t^{\prime\nu}=\left(
1,0,0,0\right)  $ ($t^{\prime\nu}$ are the components of the unit 4-vector
\emph{in the time direction in }$S^{\prime}$ and in the Einstein system of
coordinates), and consequently that $E_{f}^{\mu}$ and $B_{f}^{\mu}$ from
(\ref{gt}) transform to $E_{st.}^{\prime\mu}$ and $B_{st.}^{\prime\mu}$ in
$S^{\prime},$
\begin{align}
E_{st.}^{\prime\mu}  &  =F^{\prime\mu\nu}t_{\nu}^{\prime},B_{st.}^{\prime\mu
}=(F^{\ast})^{\prime\mu\nu}t_{\nu}^{\prime};\ E_{st.}^{\prime\mu}=\left(
0,E^{1},\gamma E^{2}-\gamma\beta B^{3},\gamma E^{3}+\gamma\beta B^{2}\right)
,\nonumber\\
B_{st.}^{\prime\mu}  &  =\left(  0,B^{1},\gamma B^{2}+\gamma\beta E^{3},\gamma
B^{3}-\gamma\beta E^{2}\right)  , \label{kr}%
\end{align}
where the subscript - st. is for - standard. \emph{The temporal components of
}$E_{st.}^{\prime\mu}$\emph{\ and }$B_{st.}^{\prime\mu}$\emph{\ in }%
$S^{\prime}$\emph{\ are again zero as are the temporal components of }%
$E_{f}^{\mu}$\emph{\ and }$B_{f}^{\mu}$\emph{\ in }$S.$\emph{\ }This fact
clearly shows that \emph{the transformations given by the relation}
(\ref{kr})\emph{\ are not the LT of some well-defined 4D quantities; the LT
cannot transform a 4-vector for which the temporal component is zero in one
frame }$S$\emph{\ to the 4-vector with the same property in relatively moving
frame }$S^{\prime}$\emph{; i.e., they cannot transform the unit 4-vector in
the time direction in one frame} $S$ \emph{to the unit 4-vector in the time
direction in another relatively moving frame }$S^{\prime}.$ Obviously
$E_{st.}^{\prime\mu}$ and $B_{st.}^{\prime\mu}$ are completely different
quantities than $E_{f}^{\prime\mu}$ and $B_{f}^{\prime\mu}$ (\ref{ebcr}) that
are obtained by the correct LT. We can easily check that
\begin{equation}
E_{st.}^{\prime\mu}\gamma_{\mu}^{\prime}\neq E_{f}^{\mu}\gamma_{\mu},\quad
B_{st.}^{\prime\mu}\gamma_{\mu}^{\prime}\neq B_{f}^{\mu}\gamma_{\mu}.
\label{krive}%
\end{equation}
This means that, e.g., $E_{f}^{\mu}\gamma_{\mu}$ and $E_{st.}^{\prime\mu
}\gamma_{\mu}^{\prime}$ \emph{are not the same quantity for observers in} $S$
\emph{and} $S^{\prime}.$ As far as relativity is concerned the quantities,
e.g., $E_{f}^{\mu}\gamma_{\mu}$ and $E_{st.}^{\prime\mu}\gamma_{\mu}^{\prime
},$ are not related to one another. The observers in $S$ and $S^{\prime}$ are
not looking at the same physical object but at two different objects;
\emph{every observer makes measurement on its own object and such measurements
are not related by the LT.} The transformations (\ref{kr}) are not the LT and
$E_{st.}^{\prime\mu}$ and $B_{st.}^{\prime\mu}$, in contrast to $E_{f}%
^{\prime\mu}$\ and $B_{f}^{\prime\mu},$\ are not well-defined 4D quantities.
From \emph{the relativistically incorrect transformations} (\ref{kr}) one
simply derives the transformations of the spatial components $E_{st.}^{\prime
i}$ and $B_{st.}^{\prime i}$, which are the same as (\ref{sk1}). It can be
again seen from (\ref{kr}), or (\ref{sk1}), \emph{that the transformations of}
$E_{st.}^{\prime i}$ \emph{and} $B_{st.}^{\prime i}$ \emph{are exactly the ST
of components of the 3-vectors} $\mathbf{E}$ \emph{and} $\mathbf{B}$ that are
obtained by Lorentz [1] and independently by Einstein [2] and subsequently
quoted in almost every textbook and paper on relativistic electrodynamics.
Notice that, in the tensor formalism as well, \emph{according to the ST}
(\ref{kr}), \emph{i.e.}, (\ref{sk1}), \emph{the transformed components}
$E_{st}^{\prime i},$ and $B_{st}^{\prime i},$ \emph{are expressed by the
mixture of components} $E_{f}^{i}$ \emph{and} $B_{f}^{i}.$ This completely
differs from the correct LT (\ref{ebcr}). Both the transformations (\ref{kr})
and the transformations for $E_{st.}^{\prime i}$ and $B_{st.}^{\prime i}$
(\ref{sk1}) are typical examples of the ''apparent'' transformations that are
first discussed in [24] and [25]. The ''apparent'' transformations of the
spatial distances (the Lorentz contraction) and the temporal distances (the
dilatation of time) are elaborated in detail in [11,12] (see also [20]). It is
explicitly shown in [12] that the true agreement with experiments that test SR
exists only when the theory deals with well-defined 4D quantities, i.e., the
quantities that are invariant upon the passive LT. In all previous treatments
of SR, e.g., [2--4], the transformations for $E_{st.}^{\prime i}$ and
$B_{st.}^{\prime i}$ (\ref{sk1}) are considered to be the LT of the 3D
electric and magnetic fields. However as shown above (the comparison of
(\ref{ebcr}) and (\ref{kr}), or (\ref{sk1})) the transformations for
$E_{st.}^{\prime i}$ and $B_{st.}^{\prime i}$ are derived from \emph{the
relativistically incorrect transformations} (\ref{kr}) and moreover the
3-vectors $\mathbf{E}^{\prime}$ and $\mathbf{B}^{\prime}$ are again formed by
an incorrect procedure in 4D spacetime, i.e., by multiplying these
\emph{relativistically incorrect components} with \emph{the unit 3-vectors in
the }$S^{\prime}$ \emph{frame.}

Let us now perform the ST of the ME (\ref{me1}) supposing that $E_{f}^{\mu}%
$\ and $B_{f}^{\mu}$\ in $S$\ are transformed into $E_{st.}^{\prime\mu}$ and
$B_{st.}^{\prime\mu}$ in $S^{\prime}$ according to (\ref{kr}) and that the set
of components $t^{\nu}=\left(  1,0,0,0\right)  $ from $S$ transforms to
$t^{\prime\nu}=\left(  1,0,0,0\right)  $ in $S^{\prime}.$ Then (\ref{me1})
transforms to the same equations but with $E_{st.}^{\prime\mu}$ and
$B_{st.}^{\prime\mu}$ replacing $E_{f}^{\mu}$\ and $B_{f}^{\mu}$\ and
$t^{\prime\nu}$ replacing $t^{\nu}$. From the transformed equations obtained
in such a way one easily finds the ST of the ME (\ref{me}). They are
\begin{align}
(\partial_{k}^{\prime}E_{st}^{\prime k}-j^{\prime0}/c\varepsilon_{0}%
)\gamma_{0}^{\prime}+(-\partial_{0}^{\prime}E_{st}^{\prime i}+c\varepsilon
^{ijk0}\partial_{j}^{\prime}B_{st,k}^{\prime}-j^{\prime i}/c\varepsilon
_{0})\gamma_{i}^{\prime}  &  =0,\nonumber\\
(-c\partial_{k}^{\prime}B_{st}^{\prime k})\gamma_{0}^{\prime}+(c\partial
_{0}^{\prime}B_{st}^{\prime i}+\varepsilon^{ijk0}\partial_{j}^{\prime}%
E_{st,k}^{\prime})\gamma_{i}^{\prime}  &  =0. \label{cm}%
\end{align}
The equations (\ref{cm}) correspond to the equation (\ref{EBC}) in the
formalism with 1-vectors $E_{f}$ and $B_{f}$. They are of the same form as the
original ME (\ref{me}) with primed quantities replacing the corresponding
unprimed ones, but, as remarked above, $E_{st.}^{\prime\mu}$ and
$B_{st.}^{\prime\mu}$ replace $E_{f}^{\mu}$\ and $B_{f}^{\mu}$\ from $S$.
Thence we get the same result as in the geometric algebra formalism, i.e.,
that \emph{the equations }(\ref{cm})\emph{\ are not the correct LT but
relativistically incorrect transformations of the original ME} (\ref{me});
\emph{the LT of the ME} (\ref{me}) \emph{are the equations} (\ref{abc})
\emph{with} (\ref{ac0}), where the Lorentz transformed electric and magnetic
fields, the components $E_{f}^{\prime\mu}$ and $B_{f}^{\prime\mu}$
respectively$,$ are given by the relations (\ref{ebcr}). We note that
Einstein's derivation [2] of the ST of fields and of the ME, together with the
similar derivation presented in [4], is already discussed in detail in [11]
and will not be repeated here. \bigskip\medskip

\textbf{D. Lorentz\ invariant\ field\ equations\ with 4-vectors }$E^{a}%
$\textbf{\ and }$B^{a}$ \bigskip

In a completely similar way as in section II.D. we perform here \emph{an
observer independent} decomposition of $F^{ab}$ into 4-vectors of the electric
$E^{a}$ and magnetic $B^{a}$ fields that are defined without reference frames,
i.e., \emph{they are independent of the chosen reference frame and of the
chosen system of coordinates in it.} (This decomposition and many results
quoted here are already presented and discussed in [11] and also in [18].)
Formally all results here can be obtained from the equations given in sections
IV.A. and IV.B. replacing in them the quantities from the rest frame of
'fiducial' observers, i.e., the $\gamma_{0}$ - frame, $t^{a},$ $E_{f}^{a}$ and
$B_{f}^{a},$ by the quantities defined without reference frames, $v^{a}$,
$E^{a}$ and $B^{a},$ respectively. Thus instead of (\ref{veef}) we have a
Lorentz invariant decomposition
\begin{align}
F^{ab}  &  =\delta_{\quad cd}^{ab}E^{c}v^{d}+c\varepsilon^{abcd}v_{c}%
B_{d},\nonumber\\
E^{a}  &  =F^{ab}v_{b},\quad B^{a}=(1/2c)\varepsilon^{abcd}v_{b}F_{cd}.
\label{cf}%
\end{align}
Inserting (\ref{cf}) into (\ref{maxten}) we find the Lorentz invariant field
equations with 4-vectors $E^{a}$ and $B^{a}$, or better to say the field
equations (with $E^{a}$ and $B^{a}$) that are defined without reference
frames
\begin{align}
(-g)^{-1/2}\partial_{a}[(-g)^{1/2}(\delta_{\quad cd}^{ab}E^{c}v^{d}%
+c\varepsilon^{abcd}v_{c}B_{d})]  &  =j^{b}/\varepsilon_{0}c,\nonumber\\
\varepsilon^{abcd}\partial_{b}[(E_{c}v_{d}-E_{d}v_{c})+c\varepsilon
_{cdef}v^{e}B^{f}]  &  =0, \label{em1}%
\end{align}
where a, b, ...., f are all the abstract indices. When writing (\ref{em1}) as
coordinate-based geometric equations in the $\left\{  \gamma_{\mu}\right\}  $
basis they become
\begin{align}
\partial_{\alpha}(\delta_{\quad\mu\nu}^{\alpha\beta}E^{\mu}v^{\nu
}+c\varepsilon^{\alpha\beta\mu\nu}v_{\mu}B_{\nu})\gamma_{\beta}  &
=(j^{\beta}/\varepsilon_{0})\gamma_{\beta}\nonumber\\
\partial_{\alpha}(\delta_{\quad\mu\nu}^{\alpha\beta}v^{\mu}cB^{\nu
}+\varepsilon^{\alpha\beta\mu\nu}v_{\mu}E_{\nu})\gamma_{\beta}  &  =0.
\label{em2}%
\end{align}
(The equations from (\ref{em2}) correspond to (\ref{me1}) but with the above
mentioned replacements.) It is clear from their form that \emph{the equations}
(\ref{em2}) \emph{are invariant upon the LT.} The usual ME (\ref{me}) are
simply obtained from (\ref{em2}) specifying that $v^{\alpha}/c=t^{\alpha}$,
i.e., choosing the rest frame of 'fiducial' observers, the $\gamma_{0}$ -
frame. In a relatively moving frame $S^{\prime}$ all quantities in (\ref{em2})
will be replaced with the primed quantities, but due to their invariance upon
the LT the equations with primed quantities are exactly equal to the
corresponding equations in $S$ (given by (\ref{em2})). Setting that
$v^{\prime\alpha}$ in the transformed (\ref{em2}) is the LT of the components
$t^{\alpha}$, i.e., $v^{\prime\alpha}=(\gamma c,\beta\gamma c,0,0)$, one
easily finds the Lorentz transformed ME (\ref{abc}) with (\ref{ac0}). Thus
both the ME (\ref{me}) and their LT (\ref{abc}) with (\ref{ac0}) are obtained
in a simple manner from (\ref{em2}).\bigskip\medskip

\textbf{V. DISCUSSION AND\ SHORT COMPARISON\ WITH\ EXPERIMENTS }\bigskip

The results obtained in this paper reveal that the usual formulation of the
relativistic electrodynamics which uses the ST of the electric and magnetic
fields and of the ME cannot be in agreement with experiments that test SR,
i.e., in which the observers from two frames of reference compare their
measurements of\emph{\ the same physical quantity.} The careful analysis of
the traditional experiments that test SR and their modern versions is reported
in [12] and it undoubtedly shows that the usual formulation of SR is only in
an ''apparent'' agreement with experiments. All usual explanations invoke the
Lorentz contraction, the dilatation of time and/or the ST of the 3D
$\mathbf{E}$ and $\mathbf{B.}$ However, as shown in [11] and [12] (see also
[20]), the Lorentz contraction (the dilatation of time) refer to the
comparison of two spatial (temporal) distances in two inertial frames of
reference, which means that they have nothing in common with the LT; the LT
cannot connect spatial (temporal) distances taken separately, see Figs. 3. and
4. in [11] for the Lorentz contraction and the dilatation of time
respectively. The essential point which is illustrated by Figs. 3. and 4. is
that, e.g., the Lorentz contracted length and the rest length do not refer to
the same quantity in the 4D spacetime. They are different quantities in the 4D
spacetime not only for different inertial frames of reference but also for
different synchronizations. Only the spacetime length does have a well-defined
physical sense in the 4D spacetime, see Figs. 1. and 2. in [11] for the
spacetime length for a moving rod and a moving clock respectively, and also
the discussion of the ''Car and garage paradox'' in the second paper in [20].

The ST of the 3D $\mathbf{E}$ and $\mathbf{B}$ are often derived, e.g., in the
well-known textbooks on electrodynamics [26], assuming the existence of the
Lorentz contraction of a moving charged system. This again shows in another
way that the ST are not relativistically correct transformations. The accepted
existence of the Lorentz contracted length of a moving object (in 1D case,
$L^{\prime}=L/\gamma$) leads many authors, e.g., [26] and [27], to the
conclusion that the charge density of a moving system of charges
($\lambda^{\prime}$) is well-defined quantity in the 4D spacetime and
consequently that it can be compared with the corresponding charge density of
the same system of charges when it is at rest ($\lambda$), $\lambda^{\prime
}=\gamma\lambda.$ Moreover the macroscopic electric charge is usually defined
both in the classical (e.g., [3], [27]) and quantum field theories (e.g.,
[28]), by the integral of the charge density over the hypersurface $t=const.,$
$Q=\int_{t=const.}\rho d^{3}x$ (in the quantum field theories $\rho=j^{0}/c$
is the charge density operator). Jackson [3], for example, explicitly argues,
when discussing the invariance of electric charge that, [3] p.549, ''the
charge in a small volume element $d^{3}x$ is $\rho d^{3}x.$ Since this is an
experimental invariant, it is true that $\rho^{\prime}d^{3}x^{\prime}=\rho
d^{3}x$.'' Thus the Lorentz contraction is always assumed in such conventional
definition. The electric charge is an experimental invariant, but it is not
correctly defined by the conventional definition. It is correctly defined as a
manifestly invariant quantity (a Lorentz scalar); the total electric charge
$Q$ in a three-dimensional hypersurface $H$ with two-dimensional boundary
$\delta H$ is defined by the tensor equation $Q_{\delta H}=(1/c)\int_{H}%
j^{a}t_{a}dH$, where $t_{a}$ is the unit normal to $H$. The charge-current
density 4-vector $j^{a}$ as a coordinate-free quantity is a well-defined 4D
quantity ($j^{a}=j^{\mu}e_{\mu}$) and not the charge density itself. All this
is discussed in much more detail in [10] and in the second paper in [20], see
also the references therein.

An important result was obtained in [20] (the second paper) using the
invariant definition of charge, particularly the fact that the charge density
is well-defined quantity in the 4D spacetime only in the rest frame of
charges. The mentioned result is that there is a second-order electric field
($\backsim v^{2}/c^{2}$, $v$ is the drift speed of the conduction charges) not
only outside a moving loop with steady current, as usually obtained (e.g.,
[27]), but also outside the same stationary loop. Of course both results refer
to superconducting loops. Namely outside a normal conductor with steady
current there is always a zero-order electric field (independent of $v$)
together with usually considered magnetic field. The results from [5,6] and
from this paper confirm in another way the mentioned results for the loop with
steady current, since the electric field as 4D quantity always transforms by
the LT again to the electric field. This means that if there is an electric
field outside a moving loop with steady current than it must exist for the
same but stationary loop. Such electric field is an experimentally verifiable
result and has to be carefully examined. The already performed experiments
[29] cannot, contrary to their claims, measure such external electric fields
(in fact, quadrupole's electric moment), but they can measure only the
potentials from monopoles. The reason is that they used probes directly
connected with superconducting wires. The experiments in [30] are better
suited for measurements of such external electric fields from steady currents
but they dealt with normal conductors and not with superconductors. The
authors of [30] forgot that always there is an external electric field for
normal current-carrying conductors. Thus their experiment actually has nothing
to do with the test of breakdown of local Lorentz invariance. However the same
type of the experiment as in [30], but with the superconducting coil, could
probably detect the external second-order electric fields. All this will be
discussed in more detail elsewhere.

Let us now briefly discuss, as an example, the Faraday disk, using both the
conventional formulation of electrodynamics with the 3D $\mathbf{E}$ and
$\mathbf{B}$ and their ST and the formulation with geometric 4D quantities,
the invariant relativistic electrodynamics (here we shall deal only with the
tensor formalism since it is better known). A conducting disk is turning about
thin axle passing through the center at a right angle to the disk and parallel
to a uniform magnetic field $\mathbf{B}$. The circuit is made by connecting
one end of the resistor to the axle (the spatial point $A$) and the other end
to a sliding contact touching the external circumference (the spatial point
$C$). The disk of radius $R$ is rotating with angular velocity $\omega.$ (For
the description and the picture of the Faraday disk see, e.g., [27] Chap. 18
or the recent paper [31].) Let us determine the electromotive force (emf) in
two inertial frames of reference, the laboratory frame $S$ in which the disk
is rotating and the frame $S^{\prime}$ instantaneously co-moving with a point
on the external circumference (say $C$, taken at some moment $t$). The
$x^{\prime}$ axis is along the velocity $\mathbf{V}$ of the point $C$ at $t$
and it is parallel to the $x$ axis. Actually all axes in $S^{\prime}$ are
parallel to the corresponding axes in $S$. The $y^{\prime}$ axis is along the
radius, i.e., along the segment $AC.$ First we calculate the emf using the
standard formulation. In the $S$ frame
\begin{equation}
emf=\oint(\mathbf{F}_{L}/q)\cdot\mathbf{dl=}\int_{AC}(F_{L,y}/q)dy\mathbf{=}%
\omega R^{2}B/2, \label{emf1}%
\end{equation}
where $\mathbf{F}_{L}$ is the Lorentz force $\mathbf{F}_{L}=q\mathbf{E+}%
q\mathbf{u}\times\mathbf{B}$, $\mathbf{E=}0$ in $S,$ $\mathbf{B}$ is along the
$+z$ axis, $q\mathbf{u}\times\mathbf{B}$ is the magnetic part of the Lorentz
force seen by the charges co-moving with the disk along the segment $AC$. The
integral along the segment $AC$ is taken at the same moment $t$. In the
$S^{\prime}$ frame the standard treatments suppose that the Lorentz force
becomes $\mathbf{F}_{L}^{\prime}=q\mathbf{E}^{\prime}\mathbf{+}q\mathbf{u}%
^{\prime}\times\mathbf{B}^{\prime}$, where the components of the 3D
$\mathbf{E}^{\prime}$\textbf{\ }and $\mathbf{B}^{\prime}$ are determined by
the ST (\ref{kr}). Thus it is argued in the standard formulation that in
$S^{\prime}$ the charges experiences the fields $\mathbf{E}^{\prime}%
=\gamma_{V}\mathbf{\beta}_{V}\times c\mathbf{B}$ and $\mathbf{B}^{\prime
}=\gamma_{V}\mathbf{B,}$ where $\mathbf{\beta}_{V}=(\mathbf{V/}c)\mathbf{i}$
and $\gamma_{V}=(1-\beta_{V}^{2})^{-1/2}$. Then only the $y^{\prime}$
component of the force $\mathbf{F}_{L}^{\prime}$ remains and it is
\begin{equation}
F_{L,y}^{\prime}=-qcB\beta_{u}/\gamma_{V}(1-\beta_{u}\beta_{V}). \label{fe1}%
\end{equation}
Notice that the same relation can be obtained from the definition of the
4-force (the components) $K^{\mu}=(\gamma_{u}\mathbf{F\cdot u}/c,\gamma
_{u}\mathbf{F})$ and its LT. This gives $\gamma_{u}^{\prime}F^{\prime2}=$
$\gamma_{u}F^{2}$ whence the same $F_{L,y}^{\prime}$ is obtained. (This
happens here accidentally since $F_{L,y}^{\prime}$ is calculated along the $y$
axis and $\mathbf{E=}0$ in $S$. Generally the expression $\mathbf{F}%
_{L}^{\prime}=q\mathbf{E}^{\prime}\mathbf{+}q\mathbf{u}^{\prime}%
\times\mathbf{B}^{\prime}$ and the expression obtained from the LT of the
4-force will not give the same result.) In $S^{\prime}$ the velocity (in units
of c) $\beta_{u}^{\prime}$ of some point on the segment $AC$ is $\beta
_{u}^{\prime}=(\beta_{u}-\beta_{V})/(1-\beta_{u}\beta_{V})$ and the
corresponding $\gamma_{u}^{\prime}$ is $\gamma_{u}^{\prime}=\gamma_{V}%
\gamma_{u}(1-\beta_{u}\beta_{V})$. The emf is again given by the integral of
$F_{L,y}^{\prime}/q$ over the common $y,y^{\prime}$ axis (along the segment
$AC,$ $\mathbf{dl}^{\prime}\mathbf{=dl}$) taken again at the same moment of
time $t$ ($y$ axis is orthogonal to the relative velocity $\mathbf{V}$)
\begin{equation}
emf^{\prime}=\int_{AC}(F_{L,y}^{\prime}/q)dy. \label{emf2}%
\end{equation}
It is clear from the expression for the emf in $S$ (\ref{emf1}) and the
corresponding one for the emf in $S^{\prime}$ (\ref{emf2}) together with
(\ref{fe1}) that these electromotive forces, in general, are not equal,
$emf\neq emf^{\prime}$. Only in the limit $\beta_{u},\beta_{V}\ll1$
$emf^{\prime}\simeq emf$. This result explicitly shows that the standard
formulation is not relativistically correct formulation.

Let us now consider the same example in the invariant relativistic
electrodynamics. In the tensor formalism the invariant Lorentz force $K^{a}$
is investigated in [11] Sec. 6.1. In terms of $F^{ab}$ it is $K^{a}%
=(q/c)F^{ab}u_{b},$ where $u^{b}$ is the 4-velocity of a charge $q$. In the
general case of an arbitrary spacetime and when $u^{a}$ is different from
$v^{a}$ (the 4-velocity of an observer who measures $E^{a}$ and $B^{a}$), i.e.
when the charge and the observer have distinct world lines, $K^{a}$ can be
written in terms of $E^{a}$ and $B^{a}$ as a sum of the $v^{a}$ - orthogonal
component, $K_{\perp}^{a}$, and $v^{a}$ - parallel component, $K_{\parallel
}^{a}$, $K^{a}=K_{\perp}^{a}+K_{\parallel}^{a}.$ $K_{\perp}^{a}$ is
\begin{equation}
K_{\perp}^{a}=(q/c^{2})\left[  \left(  v^{b}u_{b}\right)  E^{a}+c\widetilde
{\varepsilon}^{a}\!_{bc}u^{b}B^{c}\right]  \label{kaokom}%
\end{equation}
and $\widetilde{\varepsilon}_{abc}\equiv\varepsilon_{dabc}v^{d}$ is the
totally skew-symmetric Levi-Civita pseudotensor induced on the hypersurface
orthogonal to $v^{a}$, while
\begin{equation}
K_{\parallel}^{a}=(q/c^{2})\left[  \left(  E^{b}u_{b}\right)  v^{a}\right]  .
\label{kapar}%
\end{equation}
Speaking in terms of the prerelativistic notions one can say that $K_{\perp
}^{a}$ (\ref{kaokom}) plays the role of the usual Lorentz force lying on the
3D hypersurface orthogonal to $v^{a}$, while $K_{\parallel}^{a}$ (\ref{kapar})
is related to the work done by the field on the charge. However \emph{in the
invariant SR only both components together, that is, }$K^{a}$, \emph{does have
definite physical meaning and }$K^{a}$\emph{\ defines the Lorentz force both
in the theory and in experiments.} Of course $K^{a}$, $K_{\perp}^{a}$ and
$K_{\parallel}^{a}$ are all 4D quantities defined without reference frames and
the decomposition of $K^{a}$ is an observer independent decomposition. Then we
define the emf also as an invariant 4D quantity
\begin{equation}
emf=\int_{\Gamma}(K^{a}/q)dl_{a}, \label{emfi}%
\end{equation}
where $dl_{a}$ is the infinitesimal spacetime length and $\Gamma$ is the
spacetime curve. Let the observers are at rest in the $S$ frame, $v^{\mu
}=(c,0,0,0)$ whence $E^{0}=B^{0}=0$; the $S$ frame is the rest frame of
'fiducial' observers, the $\gamma_{0}$ - frame with the $\left\{  \gamma_{\mu
}\right\}  $ basis. Thus the components of the 4-vectors (in the Einstein
system of coordinates, i.e., in the $\left\{  \gamma_{\mu}\right\}  $ basis)
are $E^{\mu}=(0,0,0,0)$, $B^{\mu}=(0,0,0,B)$, $u^{\mu}=(c,u=\omega\rho,0,0)$,
$dl^{\mu}=(0,0,dl^{2}=dy,0)$. Thence $K_{\parallel}^{a}=0$, $K_{\perp}%
^{0}=K_{\perp}^{1}=K_{\perp}^{3}=0$, $K_{\perp}^{2}=quB$. When all quantities
in (\ref{emfi}) are written as CBGQs in the $S$ frame with the $\left\{
\gamma_{\mu}\right\}  $ basis we find $emf=\omega R^{2}B/2$. Since the
expression (\ref{emfi}) is independent of the chosen reference frame and of
the chosen system of coordinates in it we shall get the same result in the
relatively moving $S^{\prime}$ frame as well;
\begin{equation}
emf=\int_{\Gamma(in\ S)}(K^{\mu}/q)dl_{\mu}=\int_{\Gamma(in\ S^{\prime}%
)}(K^{\prime\mu}/q)dl_{\mu}^{\prime}=\omega R^{2}B/2. \label{ef2}%
\end{equation}
This can be checked directly performing the LT of all 4-vectors as CBGQs from
$S$ to $S^{\prime}$ including the transformation of $v^{\mu}\gamma_{\mu}$.
Obviously the approach with Lorentz invariant 4D quantities gives the
relativistically correct answer in an enough simple and transparent way.

In a like manner we could come to the same conclusion for all experiments
particularly to those that test SR. For example for the Trouton-Noble
experiment [32] (see also [33]). In the experiment they looked for the turning
motion of a charged parallel plate capacitor suspended at rest in the frame of
the earth in order to measure the earth's motion through the ether. All
explanations, which are given until now (see, e.g., [34]), for the null result
of the experiments [32] ([33]) are not relativistically correct, since they
use ill-defined quantities in the 4D spacetime; e.g., the Lorentz contraction,
the transformation equations for the usual 3D vectors $\mathbf{E}$ and
$\mathbf{B}$ and for the torque as the 3D vector, the nonelectromagnetic
forces of undefined nature, etc.. Thus, for instance, in the first paper in
[34] it is claimed: ''In particular it was seen that the potential energy of a
charge distribution changes, due to Lorentz contraction of the system, when it
is set in motion.'' Similarly in [34] two types of the ''explanations'' of the
Trouton-Noble experiment are offered; one of them is with nonelectromagnetic
forces of undefined nature, as in [34]. In both types of the ''explanations''
the Lorentz contraction is used ($d^{3}\overline{x}=\gamma d^{3}x$) and, of
course, the standard transformations of the 3D $\mathbf{E}$ and $\mathbf{B}$.
Here, it has to be noted that often, both in the classical (e.g., [27], [34])
and quantum field theories (e.g., [28]), the electromagnetic energy and
momentum are also defined, as in the standard definition of charge, by the
integrals of the energy and momentum densities over the hypersurface
$t=const..$It is then supposed that such hypersurface transforms by the LT to
the hypersurface $t^{\prime}=const.$ in a relatively moving reference frame
$S^{\prime}$, and consequently the Lorentz contraction is assumed,
$d^{3}x^{\prime}=\gamma d^{3}x$. This is relativistically incorrect since the
LT cannot transform the hypersurface $t=const.$ in $S$ to the hypersurface
$t^{\prime}=const.$ in a relatively moving $S^{\prime}.$ This is already
examined for the classical electrodynamics (the covariant formulation in the
Einstein system of coordinates) by Rohrlich [35] and using the component form
of the electric and magnetic 4-vectors $E^{\alpha}$ and $B^{\alpha}$ (the
tensor formalism) in the first paper in [20]. Recently [10] I have presented a
Lorentz invariant formulation of the relativistic electrodynamics in the
geometric algebra formalism.That formulation is exposed exclusively in terms
of the bivector field $F$, thus without using either the electric and magnetic
fields or the electromagnetic potential. There [10] the most general, observer
independent, expressions for the stress-energy vector $T(n)$ (1-vector)$,$ the
energy density $U$ (scalar), the Poynting vector $S$ and the momentum density
$g$ (1-vectors), the angular momentum density $M$ (bivector) and the Lorentz
force $K$ (1-vector) are presented and directly derived from the field
equations with $F$. Thus, e.g., the stress-energy vector $T(n)$ (which
describes the flow of energy-momentum through a hypersurface with unit normal
$n=n(x)$) is $T(n)=Un+(1/c)S$, where the energy density $U$ is
$U=-(\varepsilon_{0}/2)\left[  (F\cdot F)+2(F\cdot n)^{2}\right]  $ and the
Poynting vector $S$ is $S=-\varepsilon_{0}c\left[  (F\cdot n)\cdot F-(F\cdot
n)^{2}n\right]  $. When such invariant 4D quantities, i.e., the quantities
defined without reference frames, or the CBGQs, are used in the comparison
with experiments then, e.g., the explanation of the Trouton-Noble experiment
is very simple and natural. The values of such quantities are the same in the
rest frame of the capacitor and in the moving frame. Thus if there is no
torque (but now as a geometric, invariant, 4D quantity) in the rest frame then
the capacitor cannot appear to be rotating in a uniformly moving frame.
However we will not discuss this problem in more detail here. It will be
reported elsewhere.

We see that the general procedure in the invariant SR is the following. All
considered quantities have to be written as geometric 4D quantities, e.g., as
abstract 4D tensors, or as the Clifford multivectors, thus \emph{as quantities
which are defined without reference frames}, like in (\ref{em1}),
(\ref{emfi}), or (\ref{deb}). The physical laws expressed in terms of such
quantities automatically include the principle of relativity and there is no
need to postulate it outside the mathematical formulation of the theory. This
is a fundamental difference relative to the standard formulation [2] of the
theory of relativity. Then an appropriate reference frame and a system of
coordinates in it are chosen (in which the calculation is the simplest one)
and the quantities are written as CBGQs in that chosen system of coordinates.
The same result can be obtained in any other relatively moving inertial frame
of reference and with any permissible system of coordinates in it (including
different synchronizations) by performing the LT of all quantities (the form
of the LT that is independent of the chosen system of coordinates is given in
[11] in the tensor formalism and in [15] in the geometric algebra formalism).
It is essential for this Lorentz invariant approach that \emph{all observers
are looking at the same 4D physical quantity.} This is not the case for the
traditional approaches which caused many misconceptions and misunderstandings
of the SR. \bigskip\medskip

\textbf{VI.\ SUMMARY AND CONCLUSIONS}\bigskip

The covariance of the ME is cosidered to be one of the cornerstone of the
modern relativistic field theories, both classical and quantum. Einstein [2]
derived the ST of the 3D $\mathbf{E}$ and $\mathbf{B}$ assuming that the ME
with $\mathbf{E}$ and $\mathbf{B}$ must have the same form in all relatively
moving inertial frames of reference. In Einstein's formulation of SR [2] the
principle of relativity is a fundamental postulate that is supposed to hold
for all physical laws including those expressed by 3D quantities, e.g., the ME
with the 3D $\mathbf{E}$ and $\mathbf{B.}$ The results presented in this paper
substantially change generally accepted opinion about the covariance of the ME
exactly proving in geometric algebra and tensor formalisms that \emph{the
usual ME} ((\ref{MEC}), or (\ref{H2}), or (\ref{me})) \emph{change their form
upon the LT} (see (\ref{mcr}) with (\ref{anu}), or (\ref{ehbc}) with
(\ref{act}), or (\ref{abc}) with (\ref{ac0})). \emph{It is also proved that
the ST of the} \emph{ME} (see (\ref{rtr}) and (\ref{EBC}), or (\ref{gc}) and
(\ref{meq}), or (\ref{cm})), \emph{which leave unchanged the form of the ME,
actually have nothing in common with the LT of the usual ME.} The difference
between the LT of the ME, e.g., (\ref{mcr}) with (\ref{anu}), and their ST,
e.g., (\ref{rtr}) and (\ref{EBC}), is essentially the same as it is the
difference between the LT of the electric and magnetic fields (see (\ref{nle})
and (\ref{nlb}), or (\ref{eh}) and (\ref{Be}), or (\ref{ebcr})) and their ST
(see (\ref{ce}) and (\ref{B}), or (\ref{es}) and (\ref{bes}), or (\ref{kr})).
This last difference is proved in detail in [5] and [6] and that proof is only
briefly repeated in this paper. All this together reveals that, contrary to
the generally accepted opinion, \emph{the principle of relativity does not
hold for physical laws expressed by 3D quantities (a fundamental
achievement).} \emph{Any 3D quantity does not correctly transform upon the LT
and thus it does not have an independent physical reality in the 4D spacetime;
it is not the same quantity for relatively moving observers in the 4D
spacetime} (see also, e.g., Figs. 3. and 4. in [11], and [12]). Since the
usual ME change their form upon the LT they cannot describe in a
relativistically correct manner the experiments that test SR, i.e., the
experiments in which relatively moving observers measure \emph{the same 4D
physical quantity}. Therefore the new field equations with geometric 4D
quantities are constructed in geometric algebra formalism with 1-vectors $E$
and $B$ ((\ref{deb}) and (\ref{maeb})), and with bivectors $E_{HL}$ and
$B_{HL}$ ((\ref{Nf}) and (\ref{nh}), (\ref{D})), and also in the tensor
formalism with 4-vectors $E^{a}$ and $B^{a}$ ((\ref{em1}) and (\ref{em2}));
the Lorentz invariant field equations in the tensor formalism are already
presented in [11]. All quantities in these geometric equations are independent
of the chosen reference frame and of the chosen coordinate system in it. When
the $\gamma_{0}$ - frame with the $\left\{  \gamma_{\mu}\right\}  $ basis is
chosen, in which the observers who measure the electric and magnetic fields
are at rest, then all mentioned geometric equations become the usual ME. This
result explicitly shows that the correspondence principle is naturally
satisfied in the invariant SR. However, as seen here, \emph{the description
with 4D geometric quantities} \emph{is correct not only in the} $\gamma_{0}$ -
\emph{frame with the }$\left\{  \gamma_{\mu}\right\}  $ \emph{basis} \emph{but
in all other relatively moving frames and it holds for any permissible choice
of coordinates. We conclude from the results of this paper that geometric 4D
quantities, defined without reference frames or as CBGQs, do have an
independent physical reality and the relativistically correct physical laws
must be expressed in terms of such quantities.} The principle of relativity is
automatically satisfied with such quantities while in the standard formulation
of SR it is postulated outside the mathematical formulation of the theory. We
see that the role of the principle of relativity is substantially different in
the Einstein formulation of SR and in the invariant SR. The results of this
paper clearly support the latter one. Furthermore we note that all observer
independent quantities introduced here and the field equations written in
terms of them hold in the same form both in the flat and curved spacetimes.
The results obtained in this paper will have important and numerous
consequences in all relativistic field theories, classical and quantum. Some
of them will be soon examined. \medskip\bigskip

\textbf{REFERENCES}\bigskip

\noindent\lbrack1] H.A. Lorentz, Proceedings of the Academy of Sciences of Amsterdam,

6 (1904), in W. Perrett and G.B. Jeffery, in \textit{The Principle of Relativity}

(Dover, New York).

\noindent\lbrack2] A. Einstein, Ann. Physik. \textbf{17}, 891 (1905), tr. by
W. Perrett and G.B.

Jeffery, in \textit{The Principle of Relativity} (Dover, New York).

\noindent\lbrack3] J.D. Jackson, \textit{Classical Electrodynamics} (Wiley,
New York, 1977) 2nd edn.;

L.D. Landau and E.M. Lifshitz, \textit{The Classical Theory of Fields, }(Pergamon,

Oxford, 1979) 4th edn.;

\noindent\lbrack4] C.W. Misner, K.S.Thorne, and J.A. Wheeler,
\textit{Gravitation} (Freeman, San

Francisco, 1970).

\noindent\lbrack5] T. Ivezi\'{c}, Found. Phys. \textbf{33}, 1339 (2003); hep-th/0302188.

\noindent\lbrack6] T. Ivezi\'{c}, physics/\textit{0304085}.

\noindent\lbrack7] D. Hestenes, \textit{Space-Time Algebra }(Gordon and
Breach, New York, 1966);

\textit{Space-Time Calculus; }available at: http://modelingnts.la. asu.edu/evolution.

html; \textit{New Foundations for Classical Mechanics }(Kluwer Academic

Publishers, Dordrecht, 1999) 2nd. edn.; Am. J Phys. \textbf{71}, 691 (2003).

\noindent\lbrack8] C. Doran, and A. Lasenby, \textit{Geometric algebra for
physicists }

(Cambridge University Press, Cambridge, 2003); S. Gull, C. Doran, and

A. Lasenby, in \textit{Clifford (Geometric) Algebras with Applications to
Physics, }

\textit{Mathematics, and Engineering,} W.E. Baylis, Ed. (Birkhauser, Boston,

1997), Chs. 6-8..

\noindent\lbrack9] B. Jancewicz, \textit{Multivectors and Clifford Algebra in Electrodynamics}

(World Scientific, Singapore, 1989).

\noindent\lbrack10] T. Ivezi\'{c}, physics/\textit{0305092}.

\noindent\lbrack11] T. Ivezi\'{c}, Found. Phys. \textbf{31}, 1139 (2001).

\noindent\lbrack12] T. Ivezi\'{c}, Found. Phys. Lett. \textbf{15}, 27 (2002);
physics/0103026; physics/

0101091.

\noindent\lbrack13] D. Hestenes and G. Sobczyk, \textit{Clifford Algebra to
Geometric Calculus}

(Reidel, Dordrecht, 1984).

\noindent\lbrack14] M. Riesz, \textit{Clifford Numbers and Spinors}, Lecture
Series No. 38, The Institute

for Fluid Dynamics and Applied Mathematics, University of Maryland (1958).

\noindent\lbrack15] T. Ivezi\'{c}, hep-th/0207250; hep-ph/0205277.

\noindent\lbrack16] A. Einstein, Ann. Physik \textbf{49,} 769 (1916), tr. by
W. Perrett and G.B.

Jeffery, in \textit{The Principle of Relativity }(Dover, New York).

\noindent\lbrack17] H.N. N\'{u}\~{n}ez Y\'{e}pez, A.L. Salas Brito, and C.A.
Vargas, Revista

Mexicana de F\'{i}sica \textbf{34}, 636 (1988).

\noindent\lbrack18] T. Ivezi\'{c}, Annales de la Fondation Louis de Broglie
\textbf{27}, 287 (2002).

\noindent\lbrack19] S. Esposito, Found. Phys. \textbf{28}, 231 (1998).

\noindent\lbrack20] T. Ivezi\'{c}, Found. Phys. Lett. \textbf{12}, 105 (1999);
Found. Phys. Lett. \textbf{12},

507 (1999).

\noindent\lbrack21] In a private communication A. Lasenby suggested such form
with v

for the bivectors of the electric and magnetic fields in order to get

an analogy with my coordinate-free formulation with 1-vectors E and B.

\noindent\lbrack22] R.M. Wald, \textit{General Relativity} (The University of
Chicago Press, Chicago, 1984).

\noindent\lbrack23] J. Norton, Found. Phys.\textit{\ }\textbf{19}, 1215 (1989).

\noindent\lbrack24] F. Rohrlich, Nuovo Cimento B \textbf{45}, 76 (1966).

\noindent\lbrack25] A. Gamba, Am. J. Phys. \textbf{35}, 83 (1967).

\noindent\lbrack26] E.M. Purcell, \textit{Electricity and magnetism}, 2nd.edn.
(McGraw-Hill, New

York, 1985); R.P. Feynman, R.B. Leightonn and M. Sands, \textit{The Feynman }

\textit{lectures on physics Vol.2} (Addison-Wesley, Reading, 1964).

\noindent\lbrack27] W.K.H. Panofsky and M. Phillips, \textit{Classical
electricity and magnetism,}

2nd edn. (Addison-Wesley, Reading, Mass., 1962).

\noindent\lbrack28] J.D. Bjorken and S.D. Drell, \textit{Relativistic Quantum
Field} (McGraw-Hill,

New York, 1964); F. Mandl and G. Shaw, \textit{Quantum Field Theory} (John

Wiley \&Sons, New York, 1995); S. Weinberg, \textit{The} \textit{Quantum
Theory of}

\textit{Fields, Vol. I Foundations, }(Cambridge University Press, Cambridge,

1995); L. H. Ryder, \textit{Quantum Field Theory}, (Cambridge University Press,

Cambridge, 1985)

\noindent\lbrack29] W.F. Edwards, C.S. Kenyon and D.K. Lemon, Phys. Rev. D
\textbf{14,} 922 (1976);

D.K. Lemon, W.F. Edwards and C.S. Kenyon, Phys. Lett. A\textit{\ }\textbf{62},
105 (1992);

G.G. Shiskin, A.G. Shiskin, A.G. Smirnov, A.V. Dudarev, A.V. Barkov,

P.P. Zagnetov and Yu. M. Rybin, J. Phys. D: Appl. Phys. \textbf{35}, 497 (2002).

\noindent\lbrack30] U. Bartocci, F. Cardone and R. Mignani, Found. Phys. Lett.
\textbf{14}, 51 (2001).

\noindent\lbrack31] L. Nieves, M. Rodriguez, G. Spavieri and E. Tonni, Nuovo
Cimento B

116, 585 (2001).

\noindent\lbrack32] F.T. Trouton and H.R. Noble, Philos. Trans. R. Soc. London
Ser. A

\textbf{202}, 165 (1903).

\noindent\lbrack33] H.C. Hayden, Rev. Sci. Instrum. \textbf{65}, 788 (1994).

\noindent\lbrack34] A.K. Singal, J. Phys. A: Math. Gen. \textbf{25} 1605
(1992); Am. J. Phys. \textbf{61},

428 (1993); S. A. Teukolsky, Am. J. Phys. \textbf{64}, 1104 (1996); O.D.

Jefimenko, J. Phys. A: Math. Gen. \textbf{32,} 3755 (1999).

\noindent\lbrack35] F. Rohrlich, \textit{Classical charged particles},
(Addison-Wesley, Reading,

MA, 1965); Phys. Rev. D \textbf{25}, 3251 (1982).
\end{document}